\documentclass[conference]{IEEEtran}
\IEEEoverridecommandlockouts

\usepackage[style=ieee,
backend=biber,
minnames=1,
maxcitenames=6,
maxbibnames=3,
doi=true,
isbn=false,
url=false,
natbib=true,
date=year
]{biblatex}
\usepackage[T1]{fontenc}

\usepackage{orcidlink}
\usepackage{float}
\usepackage{amssymb}
\usepackage{todonotes}
\usepackage{hyperref}
\usepackage{graphicx}
\usepackage{lipsum}
\usepackage[most]{tcolorbox}
\usepackage{tikz}
\usetikzlibrary{calc,scopes,arrows.meta,arrows,decorations,shapes,intersections,patterns}
\usetikzlibrary{fit,fadings,automata,positioning, angles,quotes,babel,shadows,shadows.blur,shapes.misc}
\usepackage{tikz-3dplot}
\usepackage{xifthen}
\usepackage{amssymb}

\tikzset{fit margins/.style={/tikz/afit/.cd,#1,
    /tikz/.cd,
    inner xsep=\pgfkeysvalueof{/tikz/afit/left}+\pgfkeysvalueof{/tikz/afit/right},
    inner ysep=\pgfkeysvalueof{/tikz/afit/top}+\pgfkeysvalueof{/tikz/afit/bottom},
    xshift=-\pgfkeysvalueof{/tikz/afit/left}+\pgfkeysvalueof{/tikz/afit/right},
    yshift=-\pgfkeysvalueof{/tikz/afit/bottom}+\pgfkeysvalueof{/tikz/afit/top}},
    afit/.cd,left/.initial=2pt,right/.initial=2pt,bottom/.initial=2pt,top/.initial=2pt}

\definecolor{BlauHigh}{RGB}{96,150,180}
\definecolor{BlauMid}{RGB}{147,191,207}
\definecolor{BlauLow}{RGB}{189,205,214}
\definecolor{Beige}{RGB}{238,233,218}
\definecolor{GreenHigh}{RGB}{119, 191, 163}
\definecolor{GreenMid}{RGB}{152, 201, 163} %
\definecolor{GreenLow}{RGB}{164, 195, 178}
\definecolor{RedHigh}{RGB}{237, 106, 94}
\definecolor{RedMid}{RGB}{255, 142, 114}
\definecolor{RedLow}{RGB}{255, 175, 135}
\definecolor{rotx}{RGB}{128,29,26}
\definecolor{DeepGreen}{RGB}{0,64,32}
\definecolor{ClaySoft}{RGB}{225,200,185}
\definecolor{BlueFog}{RGB}{204,213,222}
\definecolor{SoftPeach}{RGB}{255,193,150}
\definecolor{stageBlue}{HTML}{2C445A}
\definecolor{ContextCol}{HTML}{c94d4d}
\definecolor{gsnExtensionColor}{HTML}{FF9A00} 
\definecolor{gsnInstantiationColor}{HTML}{09637E} 
\definecolor{gsnEliminationColor}{HTML}{A42A04}
\definecolor{gsnFillColor}{RGB}{235, 235, 235}
\definecolor{gsnBorderColor}{RGB}{0, 0, 0}
\definecolor{feedbackCol1}{HTML}{669bbc}
\definecolor{darkRed}{HTML}{780000}

\definecolor{BlueSystemCycle}{HTML}{09637E} 
\definecolor{RedArgumentCycle}{HTML}{FF9A00} 
\definecolor{BlauDark}{HTML}{09637E}
\definecolor{PillBlue}{HTML}{09637E}

\def\blockWidth{2.5cm}
\def\blockHeight{1cm}

\tikzset{
    shadow/.style={drop shadow={shadow xshift=.5ex, shadow yshift=-.5ex}}, 
    shadedBlauHigh/.style={top color=BlauHigh, bottom color=BlauHigh!90, draw=black},
    artifactLow/.style={draw=none, fill=BlauLow,shadow, minimum width=\blockWidth, minimum height=\blockHeight,align=center,text=white},
    artifactMid/.style={draw=none, fill=BlauMid,shadow, minimum width=\blockWidth, minimum height=\blockHeight, align=center,text=white},
    artifactHigh/.style={draw=none, fill=BlauHigh,shadow, minimum width=\blockWidth, minimum height=\blockHeight,align=center,text=white},
    concept/.style={draw=white,ultra thick, fill=BlauDark,shadow, minimum width=\blockWidth*1.5, minimum height=\blockHeight*1.5,align=center, font=\bfseries\large,text=white},
    context/.style={draw=none, fill=white,shadow, minimum width=\blockWidth, minimum height=\blockHeight,align=center,rounded corners=3pt, draw=BlauDark},
    evidence/.style={draw=BlauDark, fill=white,shadow, align=center,circle=2cm, minimum height=\blockHeight*.8,minimum width=\blockWidth*.8},
    areaGreen/.style={draw=black, fill=GreenLow, shadow,}, 
    areaRed/.style={draw=RedHigh, fill=RedHigh!15, shadow},
    areaBeige/.style={left color=Beige, right color=Beige!35, shadow, draw=black},
    areaBlue/.style={draw=BlauDark, fill=BlauDark!15, shadow},
    buffer/.style={shape border rotate=270,regular polygon,regular polygon sides=3,minimum height=1.5cm,fill=Beige},
    myarrow/.style={-{Triangle[width=7pt, length=5pt]},thick,BlauDark},
    aggregation/.style={{Diamond[open, width=8pt, length=16pt,fill=white]}-,thick},
    myarrowWhite/.style={-{Triangle[width=7pt, length=5pt]},thick,white},
    myarrowWhiteopen/.style={-{Triangle[open,width=7pt, length=5pt]},thick,white},
    myarrowopen/.style={-{Triangle[open, width=7pt, length=5pt]},thick,BlauDark},
    basicBlock/.style={fill=white,shadow, minimum width=\blockWidth, minimum height=\blockHeight,align=center, draw=BlauDark},
    ontologyElement/.style={fill=SoftPeach,shadow, draw=black,minimum width=\blockWidth, minimum height=\blockHeight,align=center,text=black},
    goal/.style={fill=white,shadow, minimum width=\blockWidth, minimum height=\blockHeight,align=center, draw=BlauDark},
    strategy/.style={fill=white,shadow, minimum width=\blockWidth, minimum height=\blockHeight,align=center, draw=BlauDark, trapezium,trapezium left angle =70, trapezium right angle = 110},
    connectorLabel/.style={pos=.5, font=\normalsize,text=black},
    gsn_shadow/.style={general shadow={shadow xshift=1.5pt, shadow yshift=-1.5pt, fill=black, opacity=0.3}},
    gsn_base_legend/.style={
        draw=gsnBorderColor, 
        line width=0.6pt,
        align=center, 
        gsn_shadow
    },
    gsn_goal_legend/.style={
        gsn_base_legend, rectangle,
        top color=gsnFillColor!90, bottom color=gsnFillColor!35
    },
    gsn_ExtendedGoal_legend/.style={
        gsn_base_legend, rectangle,
        top color=gsnExtensionColor!90, bottom color=gsnExtensionColor!35
    },
    gsn_InstantiatedGoal_legend/.style={
        gsn_base_legend, rectangle,
        top color=gsnInstantiationColor!50, bottom color=gsnInstantiationColor!15
    },
    undeveloped_legend/.style={
        draw=gsnBorderColor, 
        top color=gsnFillColor!90, bottom color=gsnFillColor!35, 
        shape=diamond, line width=0.6pt, gsn_shadow}
    }

\pgfdeclarelayer{background3}
\pgfdeclarelayer{background2}
\pgfdeclarelayer{background1}
\pgfdeclarelayer{main}
\pgfdeclarelayer{foreground}
\pgfsetlayers{background3,background2,background1,main,foreground}

\usepackage{scalerel,stackengine,amsmath}

\newcommand{\LegContextFlow}{\raisebox{2.35pt}{\begin{tikzpicture}[baseline=(char.base)]
    \node(char) [
        single arrow, 
        draw=darkRed, 
        line width=0.4pt,
        left color=darkRed, 
        right color=darkRed, 
        minimum height=1.8em, 
        minimum width=0.8em, 
        single arrow head extend=1.5pt, 
        single arrow tip angle=60,
        inner sep=0pt, 
        rounded corners=0.2pt,
        preaction={transform canvas={shift={(0.8pt,-0.8pt)}}, fill=black, opacity=0.2}
    ] {};
\end{tikzpicture}}}

\newcommand{\LoopInline}[2]{\raisebox{.2pt}{\tikz[baseline=(char.base)]{
    \draw[draw=#1, line width=1pt, dash pattern=on 2pt off 1.2pt] (0, 0.1) -- (1.7, 0.1);
    
    \fill[#1] (0.31, 0.1) -- ++(-0.15, 0.1) -- ++(0, -0.2) -- cycle;
    
    \fill[#1] (1.58, 0.1) -- ++(-0.15, 0.1) -- ++(0, -0.2) -- cycle;
    
    \node[draw=none, left color=#1, right color=#1!60, text=white, 
    rounded corners=3pt, inner sep=2pt, font=\sffamily\bfseries\tiny,
    preaction={transform canvas={shift={(0.3pt,-0.3pt)}}, fill=black, opacity=0.2}] (char) at (0.85, 0.1) {#2};
}}}

\usepackage[normalem]{ulem}

\newcommand\reduline{\bgroup\markoverwith{\textcolor{BlauDark}{\rule[-0.9ex]{2pt}{.3pt}}}\ULon}

\newcommand{\RequirementLine}[2]{%
\par\addvspace{.15\baselineskip}
  \noindent
  \reduline{%
  \hspace{0.3em}%
    \textcolor{BlauDark}{\itshape #1}
    \hfill%
    \mbox{%
      \textcolor{BlauDark}{%
        \boldmath$\blacktriangleright$%
      }%
      \hspace{0.3em}%
      \textcolor{BlauDark}{\bfseries R#2}%
    }%
  }\\[.2em]\noindent%
}

\newcommand{\LegendFramework}{\raisebox{-1mm}{\tikz{\node[gsn_goal_legend, minimum width=6mm, minimum height=2.5mm] {framework}}\hspace{1mm}}}

\newcommand{\LegendInstantiated}{\raisebox{-1mm}{\tikz{\node[gsn_InstantiatedGoal_legend, minimum width=6mm, minimum height=3mm] {instantiated}}\hspace{1mm}}}

\newcommand{\LegendExtended}{\raisebox{-1mm}{\tikz{\node[gsn_ExtendedGoal_legend, minimum width=6mm, minimum height=3mm] {extended}}\hspace{1mm}}}

\newcommand{\LegendUndeveloped}{\raisebox{-1mm}{\tikz{\node[undeveloped_legend, minimum size=8pt] {}}\hspace{1mm}}}

\newcommand{\LegendBlueLoop}[1]{%
    \raisebox{1pt}{%
        \begin{tikzpicture}[baseline=(char.base)]
            \node(char) [                
                draw=none,
                fill=BlueSystemCycle, 
                shape=signal, 
                signal from=west, signal to=east,
                inner sep=2pt, 
                minimum width=1.2cm, 
                minimum height=0.3cm, 
                text=white, 
                font=\sffamily\bfseries\tiny,
                text depth=0pt, 
                text height=0.7ex, 
                blur shadow={shadow blur steps=5, shadow blur radius=1pt, shadow xshift=1pt, shadow yshift=-1pt, shadow opacity=30}
            ] {#1};
        \end{tikzpicture}%
    }\hspace{2pt}%
}

\newcommand{\LegendOrangeLoop}[1]{%
    \raisebox{1pt}{%
        \begin{tikzpicture}[baseline=(char.base)]
            \node(char) [                
                draw=none,
                fill=RedArgumentCycle, 
                shape=signal, 
                signal from=west, signal to=east,
                inner sep=2pt, 
                minimum width=1.3cm, 
                minimum height=0.3cm, 
                text=white, 
                font=\sffamily\bfseries\tiny,
                text depth=0pt, 
                text height=0.7ex, 
                blur shadow={shadow blur steps=5, shadow blur radius=1pt, shadow xshift=1pt, shadow yshift=-1pt, shadow opacity=30}
            ] {#1};
        \end{tikzpicture}%
    }\hspace{2pt}%
}
\usepackage{cleveref}
\usepackage{stfloats}
\usepackage[caption=false,font=footnotesize]{subfig}
\usepackage{graphicx}
\usepackage{nicematrix}
\usepackage{ifthen}

\usepackage{color}

\definecolor{rotx}{RGB}{128,29,26}
\definecolor{DeepGreen}{RGB}{0,64,32}
\definecolor{SoftPeach}{RGB}{255,193,150}
\definecolor{ModernGrey}{RGB}{84,84,84}
\definecolor{ModernGreyBright}{RGB}{235, 235, 235}
\definecolor{Olive}{RGB}{136,140,121}
\definecolor{Brown}{RGB}{184,139,116}
\definecolor{RedArrow}{RGB}{176,0,0}
\definecolor{BlueArrow}{RGB}{0,70,140}

\crefformat{figure}{Fig.~#2#1#3}

\usepackage[most]{tcolorbox}
\tcbset{textmarker/.style={%
        enhanced,
        parbox=false,boxrule=0mm,boxsep=0mm,arc=0mm,
        outer arc=0mm,left=0mm,right=0mm,top=7pt,bottom=7pt,
        toptitle=0mm,bottomtitle=1mm,oversize}}

\newtcolorbox{errataBox}{textmarker,
    borderline west={7pt}{0pt}{rotx},
    before skip=4pt, 
    after skip=4pt,
    colback=rotx!10!white,
    colframe=rotx,
    enlarge left by=0pt,    
    left=11pt,   
    boxrule=1pt,
    right=6pt,
    width=\columnwidth,        
    box align=top,
    overlay={%
        \node[anchor=south east, xshift=0pt, yshift=0pt] at (frame.south east) 
            {\includegraphics[height=14pt]{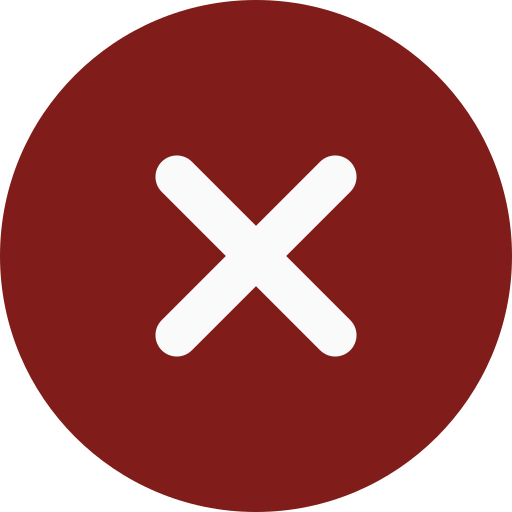}}; 
    }}             
\newtcolorbox{desiderataBox}{textmarker,
    borderline west={7pt}{0pt}{Olive},
    colframe=Olive,    
    before skip=4pt, 
    after skip=4pt,
    enlarge left by=0pt,    
    left=11pt,            
    right=6pt,
    boxrule=1pt,
    colback=Olive!10!white,
    width=\columnwidth,     
    box align=top,
    overlay={%
        \node[anchor=south east, xshift=0pt, yshift=0pt] at (frame.south east) 
            {\includegraphics[height=14pt]{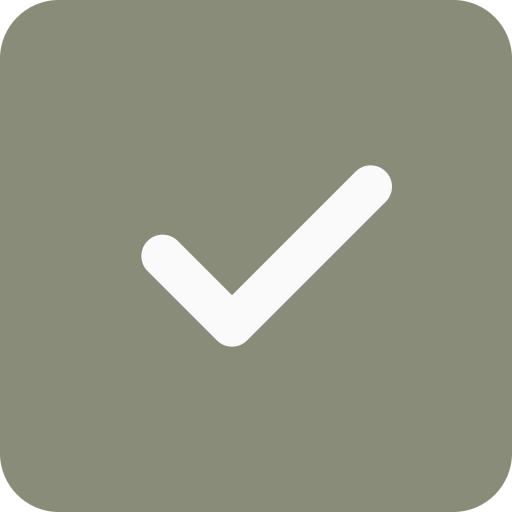}}; 
    }}

\newtcolorbox{desiderataBoxInline}[1][]{%
  enhanced,
  on line,
  box align=base,
  nobeforeafter,
  colback=Olive!10!white,
  colframe=Olive,
  borderline west={6pt}{0pt}{Olive},
  boxrule=0pt,
  left=6pt,
  width=2.4cm,
  height=.36cm,
  right=4pt,
  boxrule=1pt,
  top=0pt,
  bottom=0pt,
  sharp corners,
  valign=center,
  baseline=.0551cm, 
  #1
}

\newtcolorbox{errataBoxInline}[1][]{%
  enhanced,
  on line,
  box align=base,
  nobeforeafter,
  colback=rotx!10!white,
  borderline west={4pt}{0pt}{rotx},
    colframe=rotx,
  borderline west={6pt}{0pt}{rotx},
  boxrule=0pt,
  left=6pt,
  width=1.76cm,
  height=.36cm,
  right=4pt,
  boxrule=1pt,
  top=0pt,
  bottom=0pt,
  sharp corners,
  valign=center,
  baseline=.0551cm, 
  #1
}

\newcounter{ErrataCount}
\newcounter{DesiderataCount}

\newcommand{\ReqChecked}[1]{\raisebox{0.4ex}{$\vcenter{\hbox{\includegraphics[width=0.35cm]{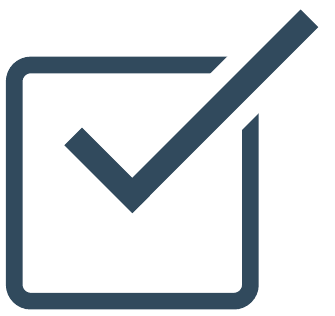}}}$}\textbf{\textcolor{BlauDark}{R#1}}}

\newcommand{\ReqUnchecked}{\raisebox{0.15ex}{$\vcenter{\hbox{\includegraphics[width=0.3cm]{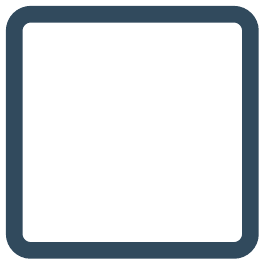}}}$}}

\newcounter{ReqCount}

\newcommand{\MacroReq}{ \refstepcounter{ReqCount}\hfill \textbf{\textcolor{BlauDark}{$\blacktriangleright$~R\theReqCount}}}

\newcommand{\ReqText}[1]{\textbf{\textcolor{BlauDark}{$\blacktriangleright$~R#1}}}

\renewcommand{\baselinestretch}{0.98}

\renewcommand*{\bibfont}{\fontsize{8}{9}\selectfont}

\usepackage[font=footnotesize, skip=3pt]{caption}

\usepackage{hologo}
\usepackage{listings}
\usepackage{xpatch}
\usepackage{xstring}

\DeclareSourcemap{
	\maps[datatype=bibtex]{
		\map{
			\step[fieldsource=langid, match=\regexp{\A(n)?((swiss)?german|austrian)\Z}, final]
			\step[fieldset=language, fieldvalue={(in German)}]
		}
		\map{
			\step[fieldsource=langid, match=pinyin, final]
			\step[fieldset=language, fieldvalue={(in Chinese)}]
		}
		\map{
			\step[fieldsource=langid, match=japanese, final]
			\step[fieldset=language, fieldvalue={(in Japanese)}]
		}
		\map{
			\step[fieldsource=langid, match=ko, final]
			\step[fieldset=language, fieldvalue={(in Korean)}]
		}
	}
}

\xpatchbibdriver{inproceedings}					 	
	{\usebibmacro{publisher+location+date}}			
	{\IfSubStr{\strfield{booktitle}}{\strfield{year}}{\usebibmacro{publisher+location}}{\usebibmacro{publisher+location+date}}} 
	{} 
	{\typeout{There was an error patching biblatex-ieee (specifically, ieee.bbx's @inproceedings driver)}} 

\newbibmacro*{publisher+location}{%
	\printlist{location}%
	\iflistundef{publisher}
	{\setunit*{\addcomma\space}}
	{\setunit*{\addcolon\space}}%
	\printlist{publisher}%
	\newunit}

\xpatchbibdriver{online}
{\printtext[parens]{\usebibmacro{date}}}
	{\iffieldundef{year}
		{}
		{\printtext[parens]{\usebibmacro{date}}}}
	{}
	{\typeout{There was an error patching biblatex-ieee (specifically, ieee.bbx's @online driver)}}

\DeclareSourcemap{
	\maps[datatype=biber]{
		\map{
			\step[fieldsource=note, final]
			\step[fieldset=addendum, origfieldval, final]
			\step[fieldset=note, null]
		}
	}
}

\DeclareBibliographyDriver{standard}{%
	\usebibmacro{bibindex}%
	\usebibmacro{begentry}%
	\usebibmacro{maintitle+title}%
	\setunit{\addcomma\addspace}%
	\usebibmacro{publisher+type+number}%
	\setunit{\labelnamepunct}\newblock
	\IfSubStr{\strfield{number}}{\strfield{year}}{}{\usebibmacro{date}} 
	\newunit\newblock
	\usebibmacro{finentry}%
}

\newbibmacro*{publisher+type+number}{%
	\printtext{%
		\printlist{publisher}
		\iflistundef{publisher}
		{\space}{}%
		\iffieldundef{type}
		{Standard}
		{\printfield{type}}
		\printfield{number}
	}%
}

\xpatchbibdriver{report}					 	
	{\usebibmacro{date}}			
	{\IfSubStr{\strfield{number}}{\strfield{year}}{}{\usebibmacro{date}}} 
	{} 
	{\typeout{There was an error patching biblatex-ieee (specifically, ieee.bbx's @report driver)}} 

\DeclareBibliographyDriver{software}{%
	\usebibmacro{bibindex}%
	\usebibmacro{begentry}%
	\usebibmacro{maintitle+title}%
	\setunit{\addcomma\addspace}%
	\usebibmacro{version+year}%
	\setunit{\labelnamepunct}\newblock
	\usebibmacro{author}%
	\setunit{\addcomma\addspace}%
	\newunit\newblock
	\usebibmacro{url+urldate}%
	\newunit\newblock
	\usebibmacro{finentry}%
}

\newbibmacro*{version+year}{%
	\printtext{%
		\iffieldundef{version}{\iffieldundef{year}{}{\printfield{year}}}{\printfield{version}}
		\printfield{number}
		\printlist{publisher}
		\iflistundef{publisher}{\space}{}%
	}%
}

\DeclareSourcemap{
	\maps[datatype=bibtex]{
		\map[overwrite, foreach={booktitle,journaltitle,eventtitle,series,publisher,institution,type}]{
			\step[fieldsource=\regexp{$MAPLOOP}, match={XXX}, replace={Acad. Serbe Sci. Arts Glas Cl. Sci. Tech.}]
			\step[fieldsource=\regexp{$MAPLOOP}, match={XXX}, replace={Acta Acust.}]
			\step[fieldsource=\regexp{$MAPLOOP}, match={XXX}, replace={Acta Astron. (Poland)}]
			\step[fieldsource=\regexp{$MAPLOOP}, match={XXX}, replace={Acta Astron. Sin. (China)}]
			\step[fieldsource=\regexp{$MAPLOOP}, match={XXX}, replace={Acta Astronaut. (U.K.)}]
			\step[fieldsource=\regexp{$MAPLOOP}, match={XXX}, replace={Acta Astrophys. Sin. (China)}]
			\step[fieldsource=\regexp{$MAPLOOP}, match={XXX}, replace={Acta Autom. Sin.}]
			\step[fieldsource=\regexp{$MAPLOOP}, match={XXX}, replace={Acta Cienc. Indica Math.}]
			\step[fieldsource=\regexp{$MAPLOOP}, match={XXX}, replace={Acta Cienc. Indica Phys.}]
			\step[fieldsource=\regexp{$MAPLOOP}, match={XXX}, replace={Acta Crystallogr. A, Found. Crystallogr.}]
			\step[fieldsource=\regexp{$MAPLOOP}, match={XXX}, replace={Acta Crystallogr. B, Struct. Sci.}]
			\step[fieldsource=\regexp{$MAPLOOP}, match={XXX}, replace={Acta Crystallogr. C, Cryst. Struct. Commun.}]
			\step[fieldsource=\regexp{$MAPLOOP}, match={XXX}, replace={Acta Electron. (France)}]
			\step[fieldsource=\regexp{$MAPLOOP}, match={XXX}, replace={Acta Cyberno}]
			\step[fieldsource=\regexp{$MAPLOOP}, match={XXX}, replace={Acta Electron. Sin. (China)}]
			\step[fieldsource=\regexp{$MAPLOOP}, match={XXX}, replace={Acta Geod. Geophys. Montan. Hung.}]
			\step[fieldsource=\regexp{$MAPLOOP}, match={XXX}, replace={Acta Geophys. Pol.}]
			\step[fieldsource=\regexp{$MAPLOOP}, match={XXX}, replace={Acta Geophys. Sin. (China)}]
			\step[fieldsource=\regexp{$MAPLOOP}, match={XXX}, replace={Acta Geophys. Sin. (USA)}]
			\step[fieldsource=\regexp{$MAPLOOP}, match={XXX}, replace={Acta Metall.}]
			\step[fieldsource=\regexp{$MAPLOOP}, match={XXX}, replace={Acta Mex. Cienc. Tecnol.}]
			\step[fieldsource=\regexp{$MAPLOOP}, match={XXX}, replace={Acta Phys. Hung.}]
			\step[fieldsource=\regexp{$MAPLOOP}, match={XXX}, replace={Acta Phys. Pol. A}]
			\step[fieldsource=\regexp{$MAPLOOP}, match={XXX}, replace={Acta Phys. Pol. B}]
			\step[fieldsource=\regexp{$MAPLOOP}, match={XXX}, replace={Acta Phys. Sin.}]
			\step[fieldsource=\regexp{$MAPLOOP}, match={XXX}, replace={Acta Phys. Slovaca}]
			\step[fieldsource=\regexp{$MAPLOOP}, match={XXX}, replace={Acta Politec. Mex.}]
			\step[fieldsource=\regexp{$MAPLOOP}, match={XXX}, replace={Acta Polytech. Scand. Appl. Phys. Ser.}]
			\step[fieldsource=\regexp{$MAPLOOP}, match={XXX}, replace={Acta Polytech. Scand. Chem. Technol.}]
			\step[fieldsource=\regexp{$MAPLOOP}, match={XXX}, replace={Metall. Ser.}]
			\step[fieldsource=\regexp{$MAPLOOP}, match={XXX}, replace={Acta Polytech. Scand. Electr. Eng. Ser.}]
			\step[fieldsource=\regexp{$MAPLOOP}, match={XXX}, replace={Acta Polytech. Scand. Math. Comput. Sci. Ser.}]
			\step[fieldsource=\regexp{$MAPLOOP}, match={XXX}, replace={Acta Polytech. Scand. Mech. Eng. Ser.}]
			\step[fieldsource=\regexp{$MAPLOOP}, match={XXX}, replace={Acta Seismol. Sin.}]
			\step[fieldsource=\regexp{$MAPLOOP}, match={XXX}, replace={Acta Tech. Acad. Sci. Hung.}]
			\step[fieldsource=\regexp{$MAPLOOP}, match={XXX}, replace={Acta Tech. CSAV}]
			\step[fieldsource=\regexp{$MAPLOOP}, match={XXX}, replace={Acustica}]
			\step[fieldsource=\regexp{$MAPLOOP}, match={XXX}, replace={(AEU) Arch. Elektr. Ubertragung}]
			\step[fieldsource=\regexp{$MAPLOOP}, match={XXX}, replace={Akust. Zh.}]
			\step[fieldsource=\regexp{$MAPLOOP}, match={XXX}, replace={Algorithmica}]
			\step[fieldsource=\regexp{$MAPLOOP}, match={XXX}, replace={Alta Freq.}]
			\step[fieldsource=\regexp{$MAPLOOP}, match={XXX}, replace={An. Acad. Bras. Cienc.}]
			\step[fieldsource=\regexp{$MAPLOOP}, match={XXX}, replace={An. Fis.}]
			\step[fieldsource=\regexp{$MAPLOOP}, match={XXX}, replace={An. Mec. Electr.}]
			\step[fieldsource=\regexp{$MAPLOOP}, match={XXX}, replace={Angew. Inform.}]
			\step[fieldsource=\regexp{$MAPLOOP}, match={XXX}, replace={Ann. Inst. Henri Poincare Phys. Theor.}]
			\step[fieldsource=\regexp{$MAPLOOP}, match={XXX}, replace={Ann. Soc. Sci. Brux. I, Sci. Math. Astron. Phys.}]
			\step[fieldsource=\regexp{$MAPLOOP}, match={XXX}, replace={Arch. Elektr. Uebertrag. (AEU)}] 
			\step[fieldsource=\regexp{$MAPLOOP}, match={XXX}, replace={Arch. Elektron. Uebertrag. Tech.}] 
			\step[fieldsource=\regexp{$MAPLOOP}, match={XXX}, replace={Arch. Elektrotech. (Poland)}]
			\step[fieldsource=\regexp{$MAPLOOP}, match={XXX}, replace={Arch. Elektrotech. (Germany)}]
			\step[fieldsource=\regexp{$MAPLOOP}, match={XXX}, replace={Ark. Fys. Semin. Trondheim}]
			\step[fieldsource=\regexp{$MAPLOOP}, match={XXX}, replace={Astrofizika}]
			\step[fieldsource=\regexp{$MAPLOOP}, match={XXX}, replace={Astron. Nachr. (Germany)}]
			\step[fieldsource=\regexp{$MAPLOOP}, match={XXX}, replace={Astron. Tidsskr.}]
			\step[fieldsource=\regexp{$MAPLOOP}, match={XXX}, replace={Astron. Vestn.}]
			\step[fieldsource=\regexp{$MAPLOOP}, match={XXX}, replace={Astron. Zh.}]
			\step[fieldsource=\regexp{$MAPLOOP}, match={XXX}, replace={Atomwirtsch.-Atomtech.}]
			\step[fieldsource=\regexp{$MAPLOOP}, match={XXX}, replace={Atti Accad. Sci. Ist. Bologna CI. Sci. Fis.}]
			\step[fieldsource=\regexp{$MAPLOOP}, match={XXX}, replace={Rend. XIII}]
			\step[fieldsource=\regexp{$MAPLOOP}, match={XXX}, replace={Atti Accad. Sci. Torino I, CI. Sci. Fis. Math. Nat.}]
			\step[fieldsource=\regexp{$MAPLOOP}, match={XXX}, replace={Autom. Strum.}]
			\step[fieldsource=\regexp{$MAPLOOP}, match={XXX}, replace={Autom. Tech. Prax.}]
			\step[fieldsource=\regexp{$MAPLOOP}, match={XXX}, replace={Automatica}]
			\step[fieldsource=\regexp{$MAPLOOP}, match={XXX}, replace={Automatie}]
			\step[fieldsource=\regexp{$MAPLOOP}, match={XXX}, replace={Automatika}]
			\step[fieldsource=\regexp{$MAPLOOP}, match={XXX}, replace={Automatisierungstechnik}]
			\step[fieldsource=\regexp{$MAPLOOP}, match={XXX}, replace={Automatizace}]
			\step[fieldsource=\regexp{$MAPLOOP}, match={XXX}, replace={Automedica}]
			\step[fieldsource=\regexp{$MAPLOOP}, match={XXX}, replace={Avtom. Telemekh.}]
			\step[fieldsource=\regexp{$MAPLOOP}, match={XXX}, replace={Avtom. Vychisl. Tekh.}]
			\step[fieldsource=\regexp{$MAPLOOP}, match={XXX}, replace={Avtomatika}]
			\step[fieldsource=\regexp{$MAPLOOP}, match={XXX}, replace={Avtometriya}]
			\step[fieldsource=\regexp{$MAPLOOP}, match={ATZelektronik worldwide}, replace={ATZelektron. worldw.}]
			\step[fieldsource=\regexp{$MAPLOOP}, match={XXX}, replace={Ber. Bunsenges. Phys. Chem.}]
			\step[fieldsource=\regexp{$MAPLOOP}, match={XXX}, replace={Biofizika}]
			\step[fieldsource=\regexp{$MAPLOOP}, match={XXX}, replace={Biometrika}]
			\step[fieldsource=\regexp{$MAPLOOP}, match={XXX}, replace={Boll. Geofis. Teor. Appl.}]
			\step[fieldsource=\regexp{$MAPLOOP}, match={XXX}, replace={Bull. Acad. Serbe Sci. Arts cl. Sci. Tech.}]
			\step[fieldsource=\regexp{$MAPLOOP}, match={XXX}, replace={Bull. Annu. Soc. Suisse Chronom. Lab. Suisse.}]
			\step[fieldsource=\regexp{$MAPLOOP}, match={XXX}, replace={Rech. Horlog.}]
			\step[fieldsource=\regexp{$MAPLOOP}, match={XXX}, replace={Bull. Cl. Sci. Acad. R. Belg.}]
			\step[fieldsource=\regexp{$MAPLOOP}, match={XXX}, replace={Bull. Dir. Etud. Rech. A}]
			\step[fieldsource=\regexp{$MAPLOOP}, match={XXX}, replace={Bull. Dir. Etud. Rech. B}]
			\step[fieldsource=\regexp{$MAPLOOP}, match={XXX}, replace={Bull. Dir. Etud. Rech. C}]
			\step[fieldsource=\regexp{$MAPLOOP}, match={XXX}, replace={Bull. Liaison Rech. Inform. Autom.}]
			\step[fieldsource=\regexp{$MAPLOOP}, match={XXX}, replace={Bur. Etud. Autom.}]
			\step[fieldsource=\regexp{$MAPLOOP}, match={XXX}, replace={CFI-Ceram. Forum Int.-Ber. Dtsch. Keram. Ges.}]
			\step[fieldsource=\regexp{$MAPLOOP}, match={XXX}, replace={Chem. Scr.}]
			\step[fieldsource=\regexp{$MAPLOOP}, match={XXX}, replace={Ciel Terre}]
			\step[fieldsource=\regexp{$MAPLOOP}, match={XXX}, replace={Cybernetica}]
			\step[fieldsource=\regexp{$MAPLOOP}, match={XXX}, replace={Deut. Hydrogr. Z.}]
			\step[fieldsource=\regexp{$MAPLOOP}, match={XXX}, replace={Dokl. Akad. Nauk SSSR}]
			\step[fieldsource=\regexp{$MAPLOOP}, match={XXX}, replace={Electroacoustique}]
			\step[fieldsource=\regexp{$MAPLOOP}, match={XXX}, replace={Electrochim. Acta}]
			\step[fieldsource=\regexp{$MAPLOOP}, match={XXX}, replace={Electrochim. Metal.}]
			\step[fieldsource=\regexp{$MAPLOOP}, match={XXX}, replace={Elektor Electron.}]
			\step[fieldsource=\regexp{$MAPLOOP}, match={XXX}, replace={Elektr. Bahnen}]
			\step[fieldsource=\regexp{$MAPLOOP}, match={XXX}, replace={Elektr. Energ.-Tech.}]
			\step[fieldsource=\regexp{$MAPLOOP}, match={XXX}, replace={Elektr. Masch.}]
			\step[fieldsource=\regexp{$MAPLOOP}, match={XXX}, replace={Elektr. Stn.}]
			\step[fieldsource=\regexp{$MAPLOOP}, match={XXX}, replace={Elektrichestvo}]
			\step[fieldsource=\regexp{$MAPLOOP}, match={XXX}, replace={Elektrie}]
			\step[fieldsource=\regexp{$MAPLOOP}, match={XXX}, replace={Elektrizitaetswirtschaft}]
			\step[fieldsource=\regexp{$MAPLOOP}, match={XXX}, replace={Elektro}]
			\step[fieldsource=\regexp{$MAPLOOP}, match={XXX}, replace={Elektro-Anz.}]
			\step[fieldsource=\regexp{$MAPLOOP}, match={XXX}, replace={Elektro-Jahr}]
			\step[fieldsource=\regexp{$MAPLOOP}, match={XXX}, replace={Elektrokhimiya}]
			\step[fieldsource=\regexp{$MAPLOOP}, match={XXX}, replace={Elektron}]
			\step[fieldsource=\regexp{$MAPLOOP}, match={XXX}, replace={Elektron Int.}]
			\step[fieldsource=\regexp{$MAPLOOP}, match={XXX}, replace={Elektron. Entwitkl.}]
			\step[fieldsource=\regexp{$MAPLOOP}, match={XXX}, replace={Elektron. Ind}]
			\step[fieldsource=\regexp{$MAPLOOP}, match={XXX}, replace={Elektron. J.}]
			\step[fieldsource=\regexp{$MAPLOOP}, match={XXX}, replace={Elektron. Prax.}]
			\step[fieldsource=\regexp{$MAPLOOP}, match={XXX}, replace={Elektron. Tekh.}]
			\step[fieldsource=\regexp{$MAPLOOP}, match={XXX}, replace={Elektronica}]
			\step[fieldsource=\regexp{$MAPLOOP}, match={XXX}, replace={Elektronik}]
			\step[fieldsource=\regexp{$MAPLOOP}, match={XXX}, replace={Elektronika}]
			\step[fieldsource=\regexp{$MAPLOOP}, match={XXX}, replace={Elektroniker}]
			\step[fieldsource=\regexp{$MAPLOOP}, match={XXX}, replace={Elektronikschau}]
			\step[fieldsource=\regexp{$MAPLOOP}, match={XXX}, replace={Elektrosvyaz}]
			\step[fieldsource=\regexp{$MAPLOOP}, match={XXX}, replace={Elektrotech. Cas.}]
			\step[fieldsource=\regexp{$MAPLOOP}, match={Elektrotechnik und Informationstechnik}, replace={Elektrotech. Inf. Tech.}]
			\step[fieldsource=\regexp{$MAPLOOP}, match={XXX}, replace={Elektrotech. Obz.}]
			\step[fieldsource=\regexp{$MAPLOOP}, match={XXX}, replace={Elektrotechniek}]
			\step[fieldsource=\regexp{$MAPLOOP}, match={XXX}, replace={Elektrotechnik (Czechoslovakia)}]
			\step[fieldsource=\regexp{$MAPLOOP}, match={XXX}, replace={Elektrotechnik (Switzerland)}]
			\step[fieldsource=\regexp{$MAPLOOP}, match={XXX}, replace={Elektrotechnik (Germany)}]
			\step[fieldsource=\regexp{$MAPLOOP}, match={XXX}, replace={Elektrotechnika}]
			\step[fieldsource=\regexp{$MAPLOOP}, match={XXX}, replace={Elektrotehnika, Zagreb}]
			\step[fieldsource=\regexp{$MAPLOOP}, match={XXX}, replace={Elektrotekhnika}]
			\step[fieldsource=\regexp{$MAPLOOP}, match={XXX}, replace={Elektroteknikeren}]
			\step[fieldsource=\regexp{$MAPLOOP}, match={XXX}, replace={Elektrowaerme Int. B.}]
			\step[fieldsource=\regexp{$MAPLOOP}, match={XXX}, replace={Elettrificazione}]
			\step[fieldsource=\regexp{$MAPLOOP}, match={XXX}, replace={Elettron. Oggi}]
			\step[fieldsource=\regexp{$MAPLOOP}, match={XXX}, replace={Elettron. Telecomun.}]
			\step[fieldsource=\regexp{$MAPLOOP}, match={XXX}, replace={Elettrotecnica}]
			\step[fieldsource=\regexp{$MAPLOOP}, match={XXX}, replace={Elek. Med Aktuell Elektron.}]
			\step[fieldsource=\regexp{$MAPLOOP}, match={XXX}, replace={Elteknik}]
			\step[fieldsource=\regexp{$MAPLOOP}, match={XXX}, replace={Energ. Atomtech.}]
			\step[fieldsource=\regexp{$MAPLOOP}, match={XXX}, replace={Energ. Elettr.}]
			\step[fieldsource=\regexp{$MAPLOOP}, match={XXX}, replace={Energetica}]
			\step[fieldsource=\regexp{$MAPLOOP}, match={XXX}, replace={Energetik}]
			\step[fieldsource=\regexp{$MAPLOOP}, match={XXX}, replace={Energetika}]
			\step[fieldsource=\regexp{$MAPLOOP}, match={XXX}, replace={Energetyka}]
			\step[fieldsource=\regexp{$MAPLOOP}, match={XXX}, replace={Energia Nuclear}]
			\step[fieldsource=\regexp{$MAPLOOP}, match={XXX}, replace={Energie Technik (Germany)}]
			\step[fieldsource=\regexp{$MAPLOOP}, match={XXX}, replace={Energie Technik (Switzerland)}]
			\step[fieldsource=\regexp{$MAPLOOP}, match={XXX}, replace={Entropie}]
			\step[fieldsource=\regexp{$MAPLOOP}, match={XXX}, replace={ETZ}]
			\step[fieldsource=\regexp{$MAPLOOP}, match={XXX}, replace={ETZ Arch.}]
			\step[fieldsource=\regexp{$MAPLOOP}, match={XXX}, replace={Feingeraetetechnik}]
			\step[fieldsource=\regexp{$MAPLOOP}, match={XXX}, replace={Feinw. Tech. Messtech.}]
			\step[fieldsource=\regexp{$MAPLOOP}, match={XXX}, replace={Fert. Tech. Betr.}]
			\step[fieldsource=\regexp{$MAPLOOP}, match={XXX}, replace={Fis. Tecnol.}]
			\step[fieldsource=\regexp{$MAPLOOP}, match={XXX}, replace={Fiz. Khim. Obrab. Mater.}]
			\step[fieldsource=\regexp{$MAPLOOP}, match={XXX}, replace={Fiz. Met. Metalloved}]
			\step[fieldsource=\regexp{$MAPLOOP}, match={XXX}, replace={Fiz. Nizk. Temp.}]
			\step[fieldsource=\regexp{$MAPLOOP}, match={XXX}, replace={Fiz. Plazmy}]
			\step[fieldsource=\regexp{$MAPLOOP}, match={XXX}, replace={Fiz. Tekh. Poluprovodn.}]
			\step[fieldsource=\regexp{$MAPLOOP}, match={XXX}, replace={Fiz. Tverd. Tela}]
			\step[fieldsource=\regexp{$MAPLOOP}, match={XXX}, replace={Fiz.-Khim. Mekh. Mater.}]
			\step[fieldsource=\regexp{$MAPLOOP}, match={XXX}, replace={Fizika}]
			\step[fieldsource=\regexp{$MAPLOOP}, match={XXX}, replace={Forsch.-Ber. Landes Nordrh.-Westfal.}]
			\step[fieldsource=\regexp{$MAPLOOP}, match={XXX}, replace={Frequenz}]
			\step[fieldsource=\regexp{$MAPLOOP}, match={XXX}, replace={Fys. Tidsskr.}]
			\step[fieldsource=\regexp{$MAPLOOP}, match={XXX}, replace={G. Fis.}]
			\step[fieldsource=\regexp{$MAPLOOP}, match={XXX}, replace={Geliotekhnika}]
			\step[fieldsource=\regexp{$MAPLOOP}, match={XXX}, replace={Geochim. Cosmochim. Acta}]
			\step[fieldsource=\regexp{$MAPLOOP}, match={XXX}, replace={Haerterei-Tech. Mitt.}]
			\step[fieldsource=\regexp{$MAPLOOP}, match={XXX}, replace={Helv. Chim. Acta}]
			\step[fieldsource=\regexp{$MAPLOOP}, match={XXX}, replace={Helv. Med. Acta}]
			\step[fieldsource=\regexp{$MAPLOOP}, match={XXX}, replace={Helv. Phys. Acta}]
			\step[fieldsource=\regexp{$MAPLOOP}, match={XXX}, replace={Hochfreq. Electroakust}]
			\step[fieldsource=\regexp{$MAPLOOP}, match={XXX}, replace={Hoppe-Seylers Z. Physiol. Chem.}]
			\step[fieldsource=\regexp{$MAPLOOP}, match={XXX}, replace={Inf. Elektron.}]
			\step[fieldsource=\regexp{$MAPLOOP}, match={XXX}, replace={Inf. Elettron.}]
			\step[fieldsource=\regexp{$MAPLOOP}, match={XXX}, replace={Inform. Forsch. Entwickl.}]
			\step[fieldsource=\regexp{$MAPLOOP}, match={XXX}, replace={Inform. Spektrum}]
			\step[fieldsource=\regexp{$MAPLOOP}, match={XXX}, replace={Inform.-Fachber.}]
			\step[fieldsource=\regexp{$MAPLOOP}, match={XXX}, replace={Informatie}]
			\step[fieldsource=\regexp{$MAPLOOP}, match={XXX}, replace={Informatik}]
			\step[fieldsource=\regexp{$MAPLOOP}, match={XXX}, replace={Informatologia Yugosl.}]
			\step[fieldsource=\regexp{$MAPLOOP}, match={XXX}, replace={Informatyka}]
			\step[fieldsource=\regexp{$MAPLOOP}, match={XXX}, replace={Infowelt}]
			\step[fieldsource=\regexp{$MAPLOOP}, match={XXX}, replace={Ing. Electr. Mec.}]
			\step[fieldsource=\regexp{$MAPLOOP}, match={XXX}, replace={Ing. Mec. Electr.}]
			\step[fieldsource=\regexp{$MAPLOOP}, match={XXX}, replace={Ing.-Arch.}]
			\step[fieldsource=\regexp{$MAPLOOP}, match={XXX}, replace={Inzh.-Fiz. Zh.}]
			\step[fieldsource=\regexp{$MAPLOOP}, match={XXX}, replace={Izmer. Tekh.}]
			\step[fieldsource=\regexp{$MAPLOOP}, match={XXX}, replace={Izv. Akad. Nauk Arm. SSR Ser. Tekh. Nauk}]
			\step[fieldsource=\regexp{$MAPLOOP}, match={XXX}, replace={Izv. Akad. Nauk SSSR Energ. Transp.}]
			\step[fieldsource=\regexp{$MAPLOOP}, match={XXX}, replace={Izv. Akad. Nauk SSSR Fiz. Atmos. Okeana}]
			\step[fieldsource=\regexp{$MAPLOOP}, match={XXX}, replace={Izv. Akad. Nauk SSSR Fiz. Zemli}]
			\step[fieldsource=\regexp{$MAPLOOP}, match={XXX}, replace={Izv. Akad. Nauk SSSR Ser. Fiz.}]
			\step[fieldsource=\regexp{$MAPLOOP}, match={XXX}, replace={Izv. Vyssh. Uchebn. Zaved. Elektromekh.}]
			\step[fieldsource=\regexp{$MAPLOOP}, match={XXX}, replace={Izv. Vyssh. Uchebn. Zaved. Radioelektron.}]
			\step[fieldsource=\regexp{$MAPLOOP}, match={XXX}, replace={Izv. Vyssh. Uchebn. Zaved. Radiofiz.}]
			\step[fieldsource=\regexp{$MAPLOOP}, match={XXX}, replace={J. Chim Phys. Phys.-Chim Biol.}]
			\step[fieldsource=\regexp{$MAPLOOP}, match={XXX}, replace={Kernenergie}]
			\step[fieldsource=\regexp{$MAPLOOP}, match={XXX}, replace={Kerntechnik}]
			\step[fieldsource=\regexp{$MAPLOOP}, match={XXX}, replace={Khim. Fiz.}]
			\step[fieldsource=\regexp{$MAPLOOP}, match={XXX}, replace={Kibern. Vychisl. Tekh.}]
			\step[fieldsource=\regexp{$MAPLOOP}, match={XXX}, replace={Kibernetika}]
			\step[fieldsource=\regexp{$MAPLOOP}, match={XXX}, replace={Kristallografiya}]
			\step[fieldsource=\regexp{$MAPLOOP}, match={XXX}, replace={Kvantovaya Elektron. Mosk.}]
			\step[fieldsource=\regexp{$MAPLOOP}, match={XXX}, replace={Kybernetes}]
			\step[fieldsource=\regexp{$MAPLOOP}, match={XXX}, replace={Kybernetika}]
			\step[fieldsource=\regexp{$MAPLOOP}, match={XXX}, replace={Med. Tek.}]
			\step[fieldsource=\regexp{$MAPLOOP}, match={XXX}, replace={Mekh. Avtom. Proizvod.}]
			\step[fieldsource=\regexp{$MAPLOOP}, match={XXX}, replace={Meres Autom.}]
			\step[fieldsource=\regexp{$MAPLOOP}, match={XXX}, replace={Mesures}]
			\step[fieldsource=\regexp{$MAPLOOP}, match={XXX}, replace={Metallofizika}]
			\step[fieldsource=\regexp{$MAPLOOP}, match={XXX}, replace={Metalloved. Term. Obrab. Met.}]
			\step[fieldsource=\regexp{$MAPLOOP}, match={XXX}, replace={Meteorol. Gidrol.}]
			\step[fieldsource=\regexp{$MAPLOOP}, match={XXX}, replace={Meteorol. Rundsch.}]
			\step[fieldsource=\regexp{$MAPLOOP}, match={XXX}, replace={Metrol. Apl.}]
			\step[fieldsource=\regexp{$MAPLOOP}, match={XXX}, replace={Metrologia}]
			\step[fieldsource=\regexp{$MAPLOOP}, match={XXX}, replace={Medel. Simul.}]
			\step[fieldsource=\regexp{$MAPLOOP}, match={XXX}, replace={Nachr. Dok.}]
			\step[fieldsource=\regexp{$MAPLOOP}, match={XXX}, replace={Nachr.tech. Elektron.}]
			\step[fieldsource=\regexp{$MAPLOOP}, match={XXX}, replace={Naturwissentchaften}]
			\step[fieldsource=\regexp{$MAPLOOP}, match={XXX}, replace={Neue Tech.}]
			\step[fieldsource=\regexp{$MAPLOOP}, match={XXX}, replace={Neue Tech. Buero}]
			\step[fieldsource=\regexp{$MAPLOOP}, match={XXX}, replace={Nukleonika}]
			\step[fieldsource=\regexp{$MAPLOOP}, match={XXX}, replace={Numer. Math.}]
			\step[fieldsource=\regexp{$MAPLOOP}, match={XXX}, replace={Nuovo Cimento A}]
			\step[fieldsource=\regexp{$MAPLOOP}, match={XXX}, replace={Nuovo Cimento B}]
			\step[fieldsource=\regexp{$MAPLOOP}, match={XXX}, replace={Nouvo Cimento C}]
			\step[fieldsource=\regexp{$MAPLOOP}, match={XXX}, replace={Nuovo Cimento D}]
			\step[fieldsource=\regexp{$MAPLOOP}, match={XXX}, replace={Okeanologiya}]
			\step[fieldsource=\regexp{$MAPLOOP}, match={XXX}, replace={Opt. Spektrosk.}]
			\step[fieldsource=\regexp{$MAPLOOP}, match={XXX}, replace={Opt.-Mekh. Prom.}]
			\step[fieldsource=\regexp{$MAPLOOP}, match={XXX}, replace={Optik}]
			\step[fieldsource=\regexp{$MAPLOOP}, match={XXX}, replace={Photogrammetria}]
			\step[fieldsource=\regexp{$MAPLOOP}, match={XXX}, replace={Photonics Spectra}]
			\step[fieldsource=\regexp{$MAPLOOP}, match={XXX}, replace={Pis’ma Astron. Zh. }]
			\step[fieldsource=\regexp{$MAPLOOP}, match={XXX}, replace={Pis’ma Zh. Eksp. Teor. Fiz. }]
			\step[fieldsource=\regexp{$MAPLOOP}, match={XXX}, replace={Pis’ma Zh. Tekh. Fiz. }]
			\step[fieldsource=\regexp{$MAPLOOP}, match={XXX}, replace={Poverkhn., Fiz. Khim. Mekh.}]
			\step[fieldsource=\regexp{$MAPLOOP}, match={XXX}, replace={Pr. Inst. Elektrotech.}]
			\step[fieldsource=\regexp{$MAPLOOP}, match={XXX}, replace={Prib. Sist. Upr.}]
			\step[fieldsource=\regexp{$MAPLOOP}, match={XXX}, replace={Prib. Tekh. Eksp.}]
			\step[fieldsource=\regexp{$MAPLOOP}, match={XXX}, replace={Prikl. Mat. Mekh.}]
			\step[fieldsource=\regexp{$MAPLOOP}, match={XXX}, replace={Prikl. Mekh.}]
			\step[fieldsource=\regexp{$MAPLOOP}, match={XXX}, replace={Probl. Kibern.}]
			\step[fieldsource=\regexp{$MAPLOOP}, match={XXX}, replace={Probl. Peredachi Inf.}]
			\step[fieldsource=\regexp{$MAPLOOP}, match={XXX}, replace={Proc. K. Ned. Akad. Wet. B, Palaeontol.}]
			\step[fieldsource=\regexp{$MAPLOOP}, match={XXX}, replace={Anthropol. }]
			\step[fieldsource=\regexp{$MAPLOOP}, match={XXX}, replace={Programmirovanie}]
			\step[fieldsource=\regexp{$MAPLOOP}, match={XXX}, replace={Prz. Elektrotech.}]
			\step[fieldsource=\regexp{$MAPLOOP}, match={XXX}, replace={Prz. Telekomun.}]
			\step[fieldsource=\regexp{$MAPLOOP}, match={XXX}, replace={PT/Elektrotech. Elektron.}]
			\step[fieldsource=\regexp{$MAPLOOP}, match={XXX}, replace={Radio Fernsehen Elektron.}]
			\step[fieldsource=\regexp{$MAPLOOP}, match={XXX}, replace={Radiotekh. Elektron.}]
			\step[fieldsource=\regexp{$MAPLOOP}, match={XXX}, replace={Radiotekhnika Mosk.}]
			\step[fieldsource=\regexp{$MAPLOOP}, match={XXX}, replace={Rev. Acad. Cienc. Zaragoza}]
			\step[fieldsource=\regexp{$MAPLOOP}, match={XXX}, replace={Rev. Electrotec. (Argentina)}]
			\step[fieldsource=\regexp{$MAPLOOP}, match={XXX}, replace={Rev. Electrotec. (Spain)}]
			\step[fieldsource=\regexp{$MAPLOOP}, match={XXX}, replace={Rev. Energ.}]
			\step[fieldsource=\regexp{$MAPLOOP}, match={XXX}, replace={Rev. Esp. Electron. Rev. Geofis.}]
			\step[fieldsource=\regexp{$MAPLOOP}, match={XXX}, replace={Ric. Autom.}]
			\step[fieldsource=\regexp{$MAPLOOP}, match={XXX}, replace={Ric. Spettrosc.}]
			\step[fieldsource=\regexp{$MAPLOOP}, match={XXX}, replace={Robotersysteme}]
			\step[fieldsource=\regexp{$MAPLOOP}, match={XXX}, replace={Rozpr. Electrotech.}]
			\step[fieldsource=\regexp{$MAPLOOP}, match={XXX}, replace={Sadhana}]
			\step[fieldsource=\regexp{$MAPLOOP}, match={XXX}, replace={Schweiz. Tech. Z.}]
			\step[fieldsource=\regexp{$MAPLOOP}, match={XXX}, replace={Scientia}]
			\step[fieldsource=\regexp{$MAPLOOP}, match={XXX}, replace={Siemens Forsch. Entwickl. Ber.}]
			\step[fieldsource=\regexp{$MAPLOOP}, match={XXX}, replace={Sist. Autom.}]
			\step[fieldsource=\regexp{$MAPLOOP}, match={XXX}, replace={Sitzungsber. Oester. Akad. Wiss. Math.-}]
			\step[fieldsource=\regexp{$MAPLOOP}, match={XXX}, replace={Naturwiss. Kl. Abt. II (Austria)}]
			\step[fieldsource=\regexp{$MAPLOOP}, match={XXX}, replace={Spectrochim. Acta A, Mol. Spectrosc.}]
			\step[fieldsource=\regexp{$MAPLOOP}, match={XXX}, replace={Spectrochim. Acta B, At. Spectrosc.}]
			\step[fieldsource=\regexp{$MAPLOOP}, match={XXX}, replace={Sprache Datenverarb.}]
			\step[fieldsource=\regexp{$MAPLOOP}, match={XXX}, replace={Stanki Instrum.}]
			\step[fieldsource=\regexp{$MAPLOOP}, match={XXX}, replace={Steklo Keram.}]
			\step[fieldsource=\regexp{$MAPLOOP}, match={XXX}, replace={Svetotekhnika  TE Int.}]
			\step[fieldsource=\regexp{$MAPLOOP}, match={XXX}, replace={Tech. Bull. Vevey}]
			\step[fieldsource=\regexp{$MAPLOOP}, match={XXX}, replace={Tech. Mitt. Krupp (Engl. Ed.)}]
			\step[fieldsource=\regexp{$MAPLOOP}, match={XXX}, replace={Tech. Mitt. PTT}]
			\step[fieldsource=\regexp{$MAPLOOP}, match={XXX}, replace={Tech. Mitt. RFZ}]
			\step[fieldsource=\regexp{$MAPLOOP}, match={XXX}, replace={Technica}]
			\step[fieldsource=\regexp{$MAPLOOP}, match={XXX}, replace={Tecnica}]
			\step[fieldsource=\regexp{$MAPLOOP}, match={XXX}, replace={Teh. Fiz.}]
			\step[fieldsource=\regexp{$MAPLOOP}, match={XXX}, replace={Tehnika}]
			\step[fieldsource=\regexp{$MAPLOOP}, match={XXX}, replace={Tekh. Elektrodin.}]
			\step[fieldsource=\regexp{$MAPLOOP}, match={XXX}, replace={Tekh. Kibern.}]
			\step[fieldsource=\regexp{$MAPLOOP}, match={XXX}, replace={Tekh. Kino Telev.}]
			\step[fieldsource=\regexp{$MAPLOOP}, match={XXX}, replace={Tekh. Misul}]
			\step[fieldsource=\regexp{$MAPLOOP}, match={XXX}, replace={Telekomunikacije}]
			\step[fieldsource=\regexp{$MAPLOOP}, match={XXX}, replace={Telektronikk}]
			\step[fieldsource=\regexp{$MAPLOOP}, match={XXX}, replace={Teleteknik}]
			\step[fieldsource=\regexp{$MAPLOOP}, match={XXX}, replace={Teor. Mat. Fiz.}]
			\step[fieldsource=\regexp{$MAPLOOP}, match={XXX}, replace={Teploeoergetika}]
			\step[fieldsource=\regexp{$MAPLOOP}, match={XXX}, replace={Teplofiz. Vvs. Temp.}]
			\step[fieldsource=\regexp{$MAPLOOP}, match={XXX}, replace={Tidskr. Dok.}]
			\step[fieldsource=\regexp{$MAPLOOP}, match={XXX}, replace={TN Nachr.}]
			\step[fieldsource=\regexp{$MAPLOOP}, match={XXX}, replace={Toute Electron.}]
			\step[fieldsource=\regexp{$MAPLOOP}, match={XXX}, replace={Tr. Inst. Teor. Astron.}]
			\step[fieldsource=\regexp{$MAPLOOP}, match={XXX}, replace={Ukr. Fiz. Zh.}]
			\step[fieldsource=\regexp{$MAPLOOP}, match={XXX}, replace={Usp. Fiz. Nauk}]
			\step[fieldsource=\regexp{$MAPLOOP}, match={XXX}, replace={Vak.-Tech.}]
			\step[fieldsource=\regexp{$MAPLOOP}, match={XXX}, replace={VDE Fachiber.}]
			\step[fieldsource=\regexp{$MAPLOOP}, match={XXX}, replace={VDI Z.}]
			\step[fieldsource=\regexp{$MAPLOOP}, match={XXX}, replace={Vestn. Mashinostr.}]
			\step[fieldsource=\regexp{$MAPLOOP}, match={XXX}, replace={Vestn. Mosk. Univ. 15, Vychisl. Mat. Kibern.}]
			\step[fieldsource=\regexp{$MAPLOOP}, match={XXX}, replace={Vestn. Mosk. Univ. 3, Fiz. Astron.}]
			\step[fieldsource=\regexp{$MAPLOOP}, match={XXX}, replace={Vesti Akad. Navuk BSSR Ser. Fiz. Energ. Navuk}]
			\step[fieldsource=\regexp{$MAPLOOP}, match={XXX}, replace={VGB Kraftwerkstech. (Ger. Ed.)}]
			\step[fieldsource=\regexp{$MAPLOOP}, match={XXX}, replace={Vide Couches Minces}]
			\step[fieldsource=\regexp{$MAPLOOP}, match={XXX}, replace={Vistas Astron.}]
			\step[fieldsource=\regexp{$MAPLOOP}, match={XXX}, replace={Vopr. At. Nauki Tekh. Ser., Fiz. Radiats.}]
			\step[fieldsource=\regexp{$MAPLOOP}, match={XXX}, replace={Povrezhdenii Radiats. Materialoved.}]
			\step[fieldsource=\regexp{$MAPLOOP}, match={XXX}, replace={Vopr. At. Nauki Tekh. Ser., Obshch. Yad. Fiz.}]
			\step[fieldsource=\regexp{$MAPLOOP}, match={XXX}, replace={Vuoto Sci. Tecnol.}]
			\step[fieldsource=\regexp{$MAPLOOP}, match={XXX}, replace={Wiss. Z. Friedrich-Schiller-Univ. Jena}]
			\step[fieldsource=\regexp{$MAPLOOP}, match={XXX}, replace={Nat.wiss. Reihe}]
			\step[fieldsource=\regexp{$MAPLOOP}, match={XXX}, replace={Wiss. Z. Karl-Marx-Univ. Leipz. Math.-}]
			\step[fieldsource=\regexp{$MAPLOOP}, match={XXX}, replace={Nat.wiss. Reihe}]
			\step[fieldsource=\regexp{$MAPLOOP}, match={XXX}, replace={Wiss. Z. Tech. Hochsch. Ilmenau}]
			\step[fieldsource=\regexp{$MAPLOOP}, match={XXX}, replace={Wiss. Z. Tech. Univ. Dresd.}]
			\step[fieldsource=\regexp{$MAPLOOP}, match={XXX}, replace={Wiss. Z. Tech. Univ. Karl-Marx-Stadt}]
			\step[fieldsource=\regexp{$MAPLOOP}, match={XXX}, replace={Yad. Fiz.}]
			\step[fieldsource=\regexp{$MAPLOOP}, match={XXX}, replace={Z. Angew. Math. Mech.}]
			\step[fieldsource=\regexp{$MAPLOOP}, match={XXX}, replace={Z. Angew. Math. Phys.}]
			\step[fieldsource=\regexp{$MAPLOOP}, match={XXX}, replace={Z. Met.kd.}]
			\step[fieldsource=\regexp{$MAPLOOP}, match={XXX}, replace={Z. Nat. Forsch. A, Phys. Phys. Chem. Kosmophys.}]
			\step[fieldsource=\regexp{$MAPLOOP}, match={XXX}, replace={Z. Oper. Res. A, Theor.}]
			\step[fieldsource=\regexp{$MAPLOOP}, match={XXX}, replace={Z. Oper. Res. B, Prax.}]
			\step[fieldsource=\regexp{$MAPLOOP}, match={XXX}, replace={Z. Phys.}]
			\step[fieldsource=\regexp{$MAPLOOP}, match={XXX}, replace={Z. Phys. A, At. Nuclei}]
			\step[fieldsource=\regexp{$MAPLOOP}, match={XXX}, replace={Z. Phys. B, Condens. Matter}]
			\step[fieldsource=\regexp{$MAPLOOP}, match={XXX}, replace={Z. Phys. C, Part Fields}]
			\step[fieldsource=\regexp{$MAPLOOP}, match={XXX}, replace={Z. Phys. Chem. Neue Folge}]
			\step[fieldsource=\regexp{$MAPLOOP}, match={XXX}, replace={Z. Phys. Chem., Leipz.}]
			\step[fieldsource=\regexp{$MAPLOOP}, match={XXX}, replace={Z. Phys. D, At. Mol. Clusters}]
			\step[fieldsource=\regexp{$MAPLOOP}, match={XXX}, replace={Zavod. Lab.}]
			\step[fieldsource=\regexp{$MAPLOOP}, match={XXX}, replace={Zh. Eksp. Teor. Fiz.}]
			\step[fieldsource=\regexp{$MAPLOOP}, match={XXX}, replace={Zh. Fiz. Khim.}]
			\step[fieldsource=\regexp{$MAPLOOP}, match={XXX}, replace={Zh. Prikl. Mekh. Tekh. Fiz.}]
			\step[fieldsource=\regexp{$MAPLOOP}, match={XXX}, replace={Zh. Prikl. Spektrosk.}]
			\step[fieldsource=\regexp{$MAPLOOP}, match={XXX}, replace={Zh. Tekh. Fiz.}]
			\step[fieldsource=\regexp{$MAPLOOP}, match={XXX}, replace={Zh. Vychisl. Mat. Mat. Fiz.}]
			\step[fieldsource=\regexp{$MAPLOOP}, match={XXX}, replace={Zisin, J. Seismol. Soc. Jpn.}] 
			\step[fieldsource=\regexp{$MAPLOOP}, match={Production}, replace={Prod.}]
			\step[fieldsource=\regexp{$MAPLOOP}, match={Reliability}, replace={Rel.}]
			\step[fieldsource=\regexp{$MAPLOOP}, match={Report}, replace={Rep.}]
			\step[fieldsource=\regexp{$MAPLOOP}, match={Semiconductor}, replace={Semicond.}]
			\step[fieldsource=\regexp{$MAPLOOP}, match={Research}, replace={Res.}]
			\step[fieldsource=\regexp{$MAPLOOP}, match={Sensing}, replace={Sens.}]
			\step[fieldsource=\regexp{$MAPLOOP}, match={Resonance}, replace={Reson.}]
			\step[fieldsource=\regexp{$MAPLOOP}, match={Series}, replace={Ser.}]
			\step[fieldsource=\regexp{$MAPLOOP}, match={Resources}, replace={Resour.}]
			\step[fieldsource=\regexp{$MAPLOOP}, match={Simulation}, replace={Simul.}]
			\step[fieldsource=\regexp{$MAPLOOP}, match={Reviews}, replace={Rev.}]
			\step[fieldsource=\regexp{$MAPLOOP}, match={Review}, replace={Rev.}]
			\step[fieldsource=\regexp{$MAPLOOP}, match={Singapore}, replace={Singap.}]
			\step[fieldsource=\regexp{$MAPLOOP}, match={Robotics}, replace={Robot.}]
			\step[fieldsource=\regexp{$MAPLOOP}, match={Sistema}, replace={Sist.}]
			\step[fieldsource=\regexp{$MAPLOOP}, match={Royal}, replace={Roy.}]
			\step[fieldsource=\regexp{$MAPLOOP}, match={Society}, replace={Soc.}]
			\step[fieldsource=\regexp{$MAPLOOP}, match={Safety}, replace={Saf.}]
			\step[fieldsource=\regexp{$MAPLOOP}, match={Sociological}, replace={Sociol.}]
			\step[fieldsource=\regexp{$MAPLOOP}, match={Satellite}, replace={Satell.}]
			\step[fieldsource=\regexp{$MAPLOOP}, match={Software}, replace={Softw.}]
			\step[fieldsource=\regexp{$MAPLOOP}, match={Scandinavian}, replace={Scand.}]
			\step[fieldsource=\regexp{$MAPLOOP}, match={Solar}, replace={Sol.}]
			\step[fieldsource=\regexp{$MAPLOOP}, match={Sciences}, replace={Sci.}]
			\step[fieldsource=\regexp{$MAPLOOP}, match={Science}, replace={Sci.}]
			\step[fieldsource=\regexp{$MAPLOOP}, match={Soviet}, replace={Sov.}]
			\step[fieldsource=\regexp{$MAPLOOP}, match={Section}, replace={Sect.}]
			\step[fieldsource=\regexp{$MAPLOOP}, match={Spectroscopy}, replace={Spectrosc.}]
			\step[fieldsource=\regexp{$MAPLOOP}, match={Security}, replace={Secur.}]
			\step[fieldsource=\regexp{$MAPLOOP}, match={Spectrum}, replace={Spectr.}]
			\step[fieldsource=\regexp{$MAPLOOP}, match={Seismology}, replace={Seismol.}]
			\step[fieldsource=\regexp{$MAPLOOP}, match={Speculations}, replace={Specul.}]
			\step[fieldsource=\regexp{$MAPLOOP}, match={Selected}, replace={Sel.}]
			\step[fieldsource=\regexp{$MAPLOOP}, match={Statistics}, replace={Statist.}]
			\step[fieldsource=\regexp{$MAPLOOP}, match={Structures}, replace={Struct.}]
			\step[fieldsource=\regexp{$MAPLOOP}, match={Structure}, replace={Struct.}]
			\step[fieldsource=\regexp{$MAPLOOP}, match={Terrestrial}, replace={Terr.}]
			\step[fieldsource=\regexp{$MAPLOOP}, match={Studies}, replace={Stud.}]
			\step[fieldsource=\regexp{$MAPLOOP}, match={Theoretical}, replace={Theor.}]
			\step[fieldsource=\regexp{$MAPLOOP}, match={Superconductivity}, replace={Supercond.}]
			\step[fieldsource=\regexp{$MAPLOOP}, match={Transactions}, replace={Trans.}]
			\step[fieldsource=\regexp{$MAPLOOP}, match={Supplement}, replace={Suppl.}]
			\step[fieldsource=\regexp{$MAPLOOP}, match={Translation}, replace={Transl.}]
			\step[fieldsource=\regexp{$MAPLOOP}, match={Surface}, replace={Surf.}]
			\step[fieldsource=\regexp{$MAPLOOP}, match={Transmission}, replace={Transmiss.}]
			\step[fieldsource=\regexp{$MAPLOOP}, match={Survey}, replace={Surv.}]
			\step[fieldsource=\regexp{$MAPLOOP}, match={Transportation}, replace={Transp.}]
			\step[fieldsource=\regexp{$MAPLOOP}, match={Sustainable}, replace={Sustain.}]
			\step[fieldsource=\regexp{$MAPLOOP}, match={Tutorials}, replace={Tut.}]
			\step[fieldsource=\regexp{$MAPLOOP}, match={Symposium}, replace={Symp.}]
			\step[fieldsource=\regexp{$MAPLOOP}, match={Ultrasonic}, replace={Ultrason.}]
			\step[fieldsource=\regexp{$MAPLOOP}, match={Systems}, replace={Syst.}]
			\step[fieldsource=\regexp{$MAPLOOP}, match={System}, replace={Syst.}]
			\step[fieldsource=\regexp{$MAPLOOP}, match={University}, replace={Univ.}]
			\step[fieldsource=\regexp{$MAPLOOP}, match={\detokenize{Universität}}, replace={Univ.}]
			\step[fieldsource=\regexp{$MAPLOOP}, match={\detokenize{Université}}, replace={Univ.}]
			\step[fieldsource=\regexp{$MAPLOOP}, match={Vacuum}, replace={Vac.}]
			\step[fieldsource=\regexp{$MAPLOOP}, match={Vehicles}, replace={Veh.}]
			\step[fieldsource=\regexp{$MAPLOOP}, match={Vehicle}, replace={Veh.}]
			\step[fieldsource=\regexp{$MAPLOOP}, match={Vehicular}, replace={Veh.}]
			\step[fieldsource=\regexp{$MAPLOOP}, match={Technology}, replace={Technol.}]
			\step[fieldsource=\regexp{$MAPLOOP}, match={Technological}, replace={Technol.}]
			\step[fieldsource=\regexp{$MAPLOOP}, match={Vibration}, replace={Vib.}]
			\step[fieldsource=\regexp{$MAPLOOP}, match={Telecommunications}, replace={Telecommun.}]
			\step[fieldsource=\regexp{$MAPLOOP}, match={Visual}, replace={Vis.}]
			\step[fieldsource=\regexp{$MAPLOOP}, match={Television}, replace={Telev.}]
			\step[fieldsource=\regexp{$MAPLOOP}, match={Welding}, replace={Weld.}]
			\step[fieldsource=\regexp{$MAPLOOP}, match={Temperature}, replace={Temp.}]
			\step[fieldsource=\regexp{$MAPLOOP}, match={Working}, replace={Work.}]
			\step[fieldsource=\regexp{$MAPLOOP}, match={Learning}, replace={Learn.}]
			\step[fieldsource=\regexp{$MAPLOOP}, match={Measurement}, replace={Meas.}]
			\step[fieldsource=\regexp{$MAPLOOP}, match={Letters}, replace={Lett.}]
			\step[fieldsource=\regexp{$MAPLOOP}, match={Letter}, replace={Lett.}]
			\step[fieldsource=\regexp{$MAPLOOP}, match={Mechanical}, replace={Mech.}]
			\step[fieldsource=\regexp{$MAPLOOP}, match={Mechanics}, replace={Mech.}]
			\step[fieldsource=\regexp{$MAPLOOP}, match={Mechanic}, replace={Mech.}]
			\step[fieldsource=\regexp{$MAPLOOP}, match={Lightwave}, replace={Lightw.}]
			\step[fieldsource=\regexp{$MAPLOOP}, match={Medical}, replace={Med.}]
			\step[fieldsource=\regexp{$MAPLOOP}, match={Logic, Logical}, replace={Log.}]
			\step[fieldsource=\regexp{$MAPLOOP}, match={Metals}, replace={Met.}]
			\step[fieldsource=\regexp{$MAPLOOP}, match={Luminescence}, replace={Lumin.}]
			\step[fieldsource=\regexp{$MAPLOOP}, match={Metallurgy}, replace={Metall.}]
			\step[fieldsource=\regexp{$MAPLOOP}, match={Machines}, replace={Mach.}]
			\step[fieldsource=\regexp{$MAPLOOP}, match={Machine}, replace={Mach.}]
			\step[fieldsource=\regexp{$MAPLOOP}, match={Meteorology}, replace={Meteorol.}]
			\step[fieldsource=\regexp{$MAPLOOP}, match={Magazine}, replace={Mag.}]
			\step[fieldsource=\regexp{$MAPLOOP}, match={Metropolitan}, replace={Metrop.}]
			\step[fieldsource=\regexp{$MAPLOOP}, match={Magnetics}, replace={Magn.}]
			\step[fieldsource=\regexp{$MAPLOOP}, match={Mexican, Mexico}, replace={Mex.}]
			\step[fieldsource=\regexp{$MAPLOOP}, match={Management}, replace={Manage.}]
			\step[fieldsource=\regexp{$MAPLOOP}, match={Microelectromechanical}, replace={Microelectromech.}]
			\step[fieldsource=\regexp{$MAPLOOP}, match={Managing}, replace={Manag.}]
			\step[fieldsource=\regexp{$MAPLOOP}, match={Microgravity}, replace={Microgr.}]
			\step[fieldsource=\regexp{$MAPLOOP}, match={Manufacturing}, replace={Manuf.}]
			\step[fieldsource=\regexp{$MAPLOOP}, match={Microscopy}, replace={Microsc.}]
			\step[fieldsource=\regexp{$MAPLOOP}, match={Marine}, replace={Mar.}]
			\step[fieldsource=\regexp{$MAPLOOP}, match={Microwaves}, replace={Microw.}]
			\step[fieldsource=\regexp{$MAPLOOP}, match={Microwave}, replace={Microw.}]
			\step[fieldsource=\regexp{$MAPLOOP}, match={Materials}, replace={Mater.}]
			\step[fieldsource=\regexp{$MAPLOOP}, match={Material}, replace={Mater.}]
			\step[fieldsource=\regexp{$MAPLOOP}, match={Military}, replace={Mil.}]
			\step[fieldsource=\regexp{$MAPLOOP}, match={Modeling}, replace={Model.}]
			\step[fieldsource=\regexp{$MAPLOOP}, match={Modelling}, replace={Model.}]
			\step[fieldsource=\regexp{$MAPLOOP}, match={Oceanic}, replace={Ocean.}]
			\step[fieldsource=\regexp{$MAPLOOP}, match={Molecular}, replace={Mol.}]
			\step[fieldsource=\regexp{$MAPLOOP}, match={Oceanography}, replace={Oceanogr.}]
			\step[fieldsource=\regexp{$MAPLOOP}, match={Monitoring}, replace={Monit.}]
			\step[fieldsource=\regexp{$MAPLOOP}, match={Occupation}, replace={Occupat.}]
			\step[fieldsource=\regexp{$MAPLOOP}, match={Multiphysics}, replace={Multiphys.}]
			\step[fieldsource=\regexp{$MAPLOOP}, match={Operational}, replace={Oper.}]
			\step[fieldsource=\regexp{$MAPLOOP}, match={Nanobioscience}, replace={Nanobiosci.}]
			\step[fieldsource=\regexp{$MAPLOOP}, match={Optical}, replace={Opt.}]
			\step[fieldsource=\regexp{$MAPLOOP}, match={Nanotechnology}, replace={Nanotechnol.}]
			\step[fieldsource=\regexp{$MAPLOOP}, match={Optics}, replace={Opt.}]
			\step[fieldsource=\regexp{$MAPLOOP}, match={National}, replace={Nat.}]
			\step[fieldsource=\regexp{$MAPLOOP}, match={Optimization}, replace={Optim.}]
			\step[fieldsource=\regexp{$MAPLOOP}, match={Naval}, replace={Nav.}]
			\step[fieldsource=\regexp{$MAPLOOP}, match={Organization}, replace={Org.}]
			\step[fieldsource=\regexp{$MAPLOOP}, match={Networking}, replace={Netw.}]
			\step[fieldsource=\regexp{$MAPLOOP}, match={Networked}, replace={Netw.}]
			\step[fieldsource=\regexp{$MAPLOOP}, match={Network}, replace={Netw.}]
			\step[fieldsource=\regexp{$MAPLOOP}, match={Packaging}, replace={Packag.}]
			\step[fieldsource=\regexp{$MAPLOOP}, match={Newsletter}, replace={Newslett.}]
			\step[fieldsource=\regexp{$MAPLOOP}, match={Particle}, replace={Part.}]
			\step[fieldsource=\regexp{$MAPLOOP}, match={Nondestructive}, replace={Nondestruct.}]
			\step[fieldsource=\regexp{$MAPLOOP}, match={Patent}, replace={Pat.}]
			\step[fieldsource=\regexp{$MAPLOOP}, match={Nuclear}, replace={Nucl.}]
			\step[fieldsource=\regexp{$MAPLOOP}, match={Performance}, replace={Perform.}]
			\step[fieldsource=\regexp{$MAPLOOP}, match={Numerical}, replace={Numer.}]
			\step[fieldsource=\regexp{$MAPLOOP}, match={Personal}, replace={Pers.}]
			\step[fieldsource=\regexp{$MAPLOOP}, match={Observations}, replace={Observ.}]
			\step[fieldsource=\regexp{$MAPLOOP}, match={Philosophical}, replace={Philos.}]
			\step[fieldsource=\regexp{$MAPLOOP}, match={Photonics}, replace={Photon.}]
			\step[fieldsource=\regexp{$MAPLOOP}, match={Productivity}, replace={Productiv.}]
			\step[fieldsource=\regexp{$MAPLOOP}, match={Photovoltaics}, replace={Photovolt.}]
			\step[fieldsource=\regexp{$MAPLOOP}, match={Programming}, replace={Program.}]
			\step[fieldsource=\regexp{$MAPLOOP}, match={Physics}, replace={Phys.}]
			\step[fieldsource=\regexp{$MAPLOOP}, match={Progress}, replace={Prog.}]
			\step[fieldsource=\regexp{$MAPLOOP}, match={Physiology}, replace={Physiol.}]
			\step[fieldsource=\regexp{$MAPLOOP}, match={Propagation}, replace={Propag.}]
			\step[fieldsource=\regexp{$MAPLOOP}, match={Planetary}, replace={Planet.}]
			\step[fieldsource=\regexp{$MAPLOOP}, match={Psychology}, replace={Psychol.}]
			\step[fieldsource=\regexp{$MAPLOOP}, match={Pneumatics}, replace={Pneum.}]
			\step[fieldsource=\regexp{$MAPLOOP}, match={Quality}, replace={Qual.}]
			\step[fieldsource=\regexp{$MAPLOOP}, match={Pollution}, replace={Pollut.}]
			\step[fieldsource=\regexp{$MAPLOOP}, match={Quarterly}, replace={Quart.}]
			\step[fieldsource=\regexp{$MAPLOOP}, match={Polymer}, replace={Polym.}]
			\step[fieldsource=\regexp{$MAPLOOP}, match={Radiation}, replace={Radiat.}]
			\step[fieldsource=\regexp{$MAPLOOP}, match={Polytechnic}, replace={Polytech.}]
			\step[fieldsource=\regexp{$MAPLOOP}, match={Radiology}, replace={Radiol.}]
			\step[fieldsource=\regexp{$MAPLOOP}, match={Practice}, replace={Pract.}]
			\step[fieldsource=\regexp{$MAPLOOP}, match={Reactor}, replace={React.}]
			\step[fieldsource=\regexp{$MAPLOOP}, match={Precision}, replace={Precis.}]
			\step[fieldsource=\regexp{$MAPLOOP}, match={Receivers}, replace={Receiv.}]
			\step[fieldsource=\regexp{$MAPLOOP}, match={Principles}, replace={Princ.}]
			\step[fieldsource=\regexp{$MAPLOOP}, match={Recognition}, replace={Recognit.}]
			\step[fieldsource=\regexp{$MAPLOOP}, match={Proceedings}, replace={Proc.}]
			\step[fieldsource=\regexp{$MAPLOOP}, match={Record}, replace={Rec.}]
			\step[fieldsource=\regexp{$MAPLOOP}, match={Processing}, replace={Process.}]
			\step[fieldsource=\regexp{$MAPLOOP}, match={Rehabilitation}, replace={Rehabil.}]
			\step[fieldsource=\regexp{$MAPLOOP}, match={Conversion}, replace={Convers.}]
			\step[fieldsource=\regexp{$MAPLOOP}, match={Digital}, replace={Digit.}]
			\step[fieldsource=\regexp{$MAPLOOP}, match={Convention}, replace={Conv.}]
			\step[fieldsource=\regexp{$MAPLOOP}, match={Disclosure}, replace={Discl.}]
			\step[fieldsource=\regexp{$MAPLOOP}, match={Correspondence}, replace={Corresp.}]
			\step[fieldsource=\regexp{$MAPLOOP}, match={Discussions}, replace={Discuss.}]
			\step[fieldsource=\regexp{$MAPLOOP}, match={Critical}, replace={Crit.}]
			\step[fieldsource=\regexp{$MAPLOOP}, match={Dissertations}, replace={Diss.}]
			\step[fieldsource=\regexp{$MAPLOOP}, match={Crystal}, replace={Cryst.}]
			\step[fieldsource=\regexp{$MAPLOOP}, match={Distributed}, replace={Distrib.}]
			\step[fieldsource=\regexp{$MAPLOOP}, match={Crystallography}, replace={Crystallogr.}]
			\step[fieldsource=\regexp{$MAPLOOP}, match={Dynamics}, replace={Dyn.}]
			\step[fieldsource=\regexp{$MAPLOOP}, match={Cybernetics}, replace={Cybern.}]
			\step[fieldsource=\regexp{$MAPLOOP}, match={Earthquake}, replace={Earthq.}]
			\step[fieldsource=\regexp{$MAPLOOP}, match={Decision}, replace={Decis.}]
			\step[fieldsource=\regexp{$MAPLOOP}, match={Economics}, replace={Econ.}]
			\step[fieldsource=\regexp{$MAPLOOP}, match={Economic}, replace={Econ.}]
			\step[fieldsource=\regexp{$MAPLOOP}, match={Economical}, replace={Econ.}]
			\step[fieldsource=\regexp{$MAPLOOP}, match={Edition}, replace={Ed.}]
			\step[fieldsource=\regexp{$MAPLOOP}, match={Evolutionary}, replace={Evol.}]
			\step[fieldsource=\regexp{$MAPLOOP}, match={Education}, replace={Educ.}]
			\step[fieldsource=\regexp{$MAPLOOP}, match={Exhibition}, replace={Exhib.}]
			\step[fieldsource=\regexp{$MAPLOOP}, match={Electrical}, replace={Elect.}]
			\step[fieldsource=\regexp{$MAPLOOP}, match={Electric}, replace={Elect.}]
			\step[fieldsource=\regexp{$MAPLOOP}, match={Experimental}, replace={Exp.}]
			\step[fieldsource=\regexp{$MAPLOOP}, match={Electrification}, replace={Electrific.}]
			\step[fieldsource=\regexp{$MAPLOOP}, match={Exploratory}, replace={Explor.}]
			\step[fieldsource=\regexp{$MAPLOOP}, match={Electromagnetic}, replace={Electromagn.}]
			\step[fieldsource=\regexp{$MAPLOOP}, match={Exposition}, replace={Expo.}]
			\step[fieldsource=\regexp{$MAPLOOP}, match={Electroacoustic}, replace={Electroacoust.}]
			\step[fieldsource=\regexp{$MAPLOOP}, match={Express}, replace={Express}]
			\step[fieldsource=\regexp{$MAPLOOP}, match={Electronics}, replace={Electron.}]
			\step[fieldsource=\regexp{$MAPLOOP}, match={Electronic}, replace={Electron.}]
			\step[fieldsource=\regexp{$MAPLOOP}, match={Fabrication}, replace={Fabr.}]
			\step[fieldsource=\regexp{$MAPLOOP}, match={Emerging}, replace={Emerg.}]
			\step[fieldsource=\regexp{$MAPLOOP}, match={Faculty}, replace={Fac.}]
			\step[fieldsource=\regexp{$MAPLOOP}, match={Engineering}, replace={Eng.}]
			\step[fieldsource=\regexp{$MAPLOOP}, match={Engineers}, replace={Eng.}]
			\step[fieldsource=\regexp{$MAPLOOP}, match={Engineer}, replace={Eng.}]
			\step[fieldsource=\regexp{$MAPLOOP}, match={Ferroelectrics}, replace={Ferroelect.}]
			\step[fieldsource=\regexp{$MAPLOOP}, match={Environment}, replace={Environ.}]
			\step[fieldsource=\regexp{$MAPLOOP}, match={Francais, French}, replace={Fr.}]
			\step[fieldsource=\regexp{$MAPLOOP}, match={Equations}, replace={Equ.}]
			\step[fieldsource=\regexp{$MAPLOOP}, match={Frequency}, replace={Freq.}]
			\step[fieldsource=\regexp{$MAPLOOP}, match={Equipment}, replace={Equip.}]
			\step[fieldsource=\regexp{$MAPLOOP}, match={Foundation}, replace={Found.}]
			\step[fieldsource=\regexp{$MAPLOOP}, match={Ergonomics}, replace={Ergonom.}]
			\step[fieldsource=\regexp{$MAPLOOP}, match={Fundamental}, replace={Fundam.}]
			\step[fieldsource=\regexp{$MAPLOOP}, match={European}, replace={Eur.}]
			\step[fieldsource=\regexp{$MAPLOOP}, match={Generation}, replace={Gener.}]
			\step[fieldsource=\regexp{$MAPLOOP}, match={Evaluation}, replace={Eval.}]
			\step[fieldsource=\regexp{$MAPLOOP}, match={Geology}, replace={Geol.}]
			\step[fieldsource=\regexp{$MAPLOOP}, match={Geophysics}, replace={Geophys.}]
			\step[fieldsource=\regexp{$MAPLOOP}, match={Innovations}, replace={Innov.}]
			\step[fieldsource=\regexp{$MAPLOOP}, match={Innovation}, replace={Innov.}]
			\step[fieldsource=\regexp{$MAPLOOP}, match={Geoscience}, replace={Geosci.}]
			\step[fieldsource=\regexp{$MAPLOOP}, match={Institute}, replace={Inst.}]
			\step[fieldsource=\regexp{$MAPLOOP}, match={Institution}, replace={Inst.}]
			\step[fieldsource=\regexp{$MAPLOOP}, match={Graphics}, replace={Graph.}]
			\step[fieldsource=\regexp{$MAPLOOP}, match={Instrument}, replace={Instrum.}]
			\step[fieldsource=\regexp{$MAPLOOP}, match={Guidance}, replace={Guid.}]
			\step[fieldsource=\regexp{$MAPLOOP}, match={Instrumentation}, replace={Instrum.}]
			\step[fieldsource=\regexp{$MAPLOOP}, match={Harmonics}, replace={Harmon.}]
			\step[fieldsource=\regexp{$MAPLOOP}, match={Harmonic}, replace={Harmon.}]
			\step[fieldsource=\regexp{$MAPLOOP}, match={Insulation}, replace={Insul.}]
			\step[fieldsource=\regexp{$MAPLOOP}, match={History}, replace={Hist.}]
			\step[fieldsource=\regexp{$MAPLOOP}, match={Integrated}, replace={Integr.}]
			\step[fieldsource=\regexp{$MAPLOOP}, match={Horizon}, replace={Horiz.}]
			\step[fieldsource=\regexp{$MAPLOOP}, match={Intelligence}, replace={Intell.}]
			\step[fieldsource=\regexp{$MAPLOOP}, match={Hungary}, replace={Hung.}]
			\step[fieldsource=\regexp{$MAPLOOP}, match={Hungarian}, replace={Hung.}]
			\step[fieldsource=\regexp{$MAPLOOP}, match={Intelligent}, replace={Intell.}]
			\step[fieldsource=\regexp{$MAPLOOP}, match={Hydraulics}, replace={Hydraul.}]
			\step[fieldsource=\regexp{$MAPLOOP}, match={Interactions}, replace={Interact.}]
			\step[fieldsource=\regexp{$MAPLOOP}, match={Hydrology}, replace={Hydrol.}]
			\step[fieldsource=\regexp{$MAPLOOP}, match={Internationales}, replace={Int.}]
			\step[fieldsource=\regexp{$MAPLOOP}, match={International}, replace={Int.}]
			\step[fieldsource=\regexp{$MAPLOOP}, match={Illuminating}, replace={Illum.}]
			\step[fieldsource=\regexp{$MAPLOOP}, match={Isotopes}, replace={Isot.}]
			\step[fieldsource=\regexp{$MAPLOOP}, match={Imaging}, replace={Imag.}]
			\step[fieldsource=\regexp{$MAPLOOP}, match={Israel}, replace={Isr.}]
			\step[fieldsource=\regexp{$MAPLOOP}, match={Industrial}, replace={Ind.}]
			\step[fieldsource=\regexp{$MAPLOOP}, match={Japan}, replace={Jpn.}]
			\step[fieldsource=\regexp{$MAPLOOP}, match={Information}, replace={Inf.}]
			\step[fieldsource=\regexp{$MAPLOOP}, match={Journal}, replace={J.}]
			\step[fieldsource=\regexp{$MAPLOOP}, match={Informatics}, replace={Inform.}]
			\step[fieldsource=\regexp{$MAPLOOP}, match={Knowledge}, replace={Knowl.}]
			\step[fieldsource=\regexp{$MAPLOOP}, match={Laboratory}, replace={Lab.}]
			\step[fieldsource=\regexp{$MAPLOOP}, match={Laboratories}, replace={Lab.}]
			\step[fieldsource=\regexp{$MAPLOOP}, match={Mathematical}, replace={Math.}]
			\step[fieldsource=\regexp{$MAPLOOP}, match={Language}, replace={Lang.}]
			\step[fieldsource=\regexp{$MAPLOOP}, match={Mathematics}, replace={Math.}]
			\step[fieldsource=\regexp{$MAPLOOP}, match={Abstracts}, replace={Abstr.}]
			\step[fieldsource=\regexp{$MAPLOOP}, match={Analysis}, replace={Anal.}]
			\step[fieldsource=\regexp{$MAPLOOP}, match={Academy}, replace={Acad.}]
			\step[fieldsource=\regexp{$MAPLOOP}, match={Annals}, replace={Ann.}]
			\step[fieldsource=\regexp{$MAPLOOP}, match={Accelerator}, replace={Accel.}]
			\step[fieldsource=\regexp{$MAPLOOP}, match={Annual}, replace={Annu.}]
			\step[fieldsource=\regexp{$MAPLOOP}, match={Acoustics}, replace={Acoust.}]
			\step[fieldsource=\regexp{$MAPLOOP}, match={Apparatus}, replace={App.}]
			\step[fieldsource=\regexp{$MAPLOOP}, match={Active}, replace={Act.}]
			\step[fieldsource=\regexp{$MAPLOOP}, match={Applications}, replace={Appl.}]
			\step[fieldsource=\regexp{$MAPLOOP}, match={Administration}, replace={Admin.}]
			\step[fieldsource=\regexp{$MAPLOOP}, match={Applied}, replace={Appl.}]
			\step[fieldsource=\regexp{$MAPLOOP}, match={Administrative}, replace={Administ.}]
			\step[fieldsource=\regexp{$MAPLOOP}, match={Approximate}, replace={Approx.}]
			\step[fieldsource=\regexp{$MAPLOOP}, match={Advanced}, replace={Adv.}]
			\step[fieldsource=\regexp{$MAPLOOP}, match={Advances}, replace={Adv.}]
			\step[fieldsource=\regexp{$MAPLOOP}, match={Archives}, replace={Arch.}]
			\step[fieldsource=\regexp{$MAPLOOP}, match={Archive}, replace={Arch.}]
			\step[fieldsource=\regexp{$MAPLOOP}, match={Aeronautics}, replace={Aeronaut.}]
			\step[fieldsource=\regexp{$MAPLOOP}, match={Artificial}, replace={Artif.}]
			\step[fieldsource=\regexp{$MAPLOOP}, match={Aerospace}, replace={Aerosp.}]
			\step[fieldsource=\regexp{$MAPLOOP}, match={Assembly}, replace={Assem.}]
			\step[fieldsource=\regexp{$MAPLOOP}, match={Affective}, replace={Affect.}]
			\step[fieldsource=\regexp{$MAPLOOP}, match={Association}, replace={Assoc.}]
			\step[fieldsource=\regexp{$MAPLOOP}, match={Africa}, replace={Afr.}]
			\step[fieldsource=\regexp{$MAPLOOP}, match={African}, replace={Afr.}]
			\step[fieldsource=\regexp{$MAPLOOP}, match={Astronomy}, replace={Astron.}]
			\step[fieldsource=\regexp{$MAPLOOP}, match={Aircraft}, replace={Aircr.}]
			\step[fieldsource=\regexp{$MAPLOOP}, match={Astronautics}, replace={Astronaut.}]
			\step[fieldsource=\regexp{$MAPLOOP}, match={Algebraic}, replace={Algebr.}]
			\step[fieldsource=\regexp{$MAPLOOP}, match={Astrophysics}, replace={Astrophys.}]
			\step[fieldsource=\regexp{$MAPLOOP}, match={American}, replace={Amer.}]
			\step[fieldsource=\regexp{$MAPLOOP}, match={Atmosphere}, replace={Atmos.}]
			\step[fieldsource=\regexp{$MAPLOOP}, match={Atomic}, replace={At.}]
			\step[fieldsource=\regexp{$MAPLOOP}, match={Atoms}, replace={At.}]
			\step[fieldsource=\regexp{$MAPLOOP}, match={Broadcasting}, replace={Broadcast.}]
			\step[fieldsource=\regexp{$MAPLOOP}, match={Australasian}, replace={Australas.}]
			\step[fieldsource=\regexp{$MAPLOOP}, match={Bulletin}, replace={Bull.}]
			\step[fieldsource=\regexp{$MAPLOOP}, match={Australia}, replace={Aust.}]
			\step[fieldsource=\regexp{$MAPLOOP}, match={Bureau}, replace={Bur.}]
			\step[fieldsource=\regexp{$MAPLOOP}, match={{Automatic }}, replace={Autom.}]
			\step[fieldsource=\regexp{$MAPLOOP}, match={Business}, replace={Bus.}]
			\step[fieldsource=\regexp{$MAPLOOP}, match={Automation}, replace={Automat.}]
			\step[fieldsource=\regexp{$MAPLOOP}, match={Canadian}, replace={Can.}]
			\step[fieldsource=\regexp{$MAPLOOP}, match={Automotive}, replace={Automot.}]
			\step[fieldsource=\regexp{$MAPLOOP}, match={Automobiles}, replace={Automob.}]
			\step[fieldsource=\regexp{$MAPLOOP}, match={Automobile}, replace={Automob.}]
			\step[fieldsource=\regexp{$MAPLOOP}, match={Ceramic}, replace={Ceram.}]
			\step[fieldsource=\regexp{$MAPLOOP}, match={Chemical}, replace={Chem.}]
			\step[fieldsource=\regexp{$MAPLOOP}, match={Behavioral}, replace={Behav.}]
			\step[fieldsource=\regexp{$MAPLOOP}, match={Behavior}, replace={Behav.}]
			\step[fieldsource=\regexp{$MAPLOOP}, match={Chinese}, replace={Chin.}]
			\step[fieldsource=\regexp{$MAPLOOP}, match={Belgian}, replace={Belg.}]
			\step[fieldsource=\regexp{$MAPLOOP}, match={Climatology}, replace={Climatol.}]
			\step[fieldsource=\regexp{$MAPLOOP}, match={Biochemical}, replace={Biochem.}]
			\step[fieldsource=\regexp{$MAPLOOP}, match={Clinical}, replace={Clin.}]
			\step[fieldsource=\regexp{$MAPLOOP}, match={Bioinformatics}, replace={Bioinf.}]
			\step[fieldsource=\regexp{$MAPLOOP}, match={Cognitive}, replace={Cogn.}]
			\step[fieldsource=\regexp{$MAPLOOP}, match={Biology, Biological}, replace={Biol.}]
			\step[fieldsource=\regexp{$MAPLOOP}, match={Colloquium}, replace={Colloq.}]
			\step[fieldsource=\regexp{$MAPLOOP}, match={Kolloquium}, replace={Kolloq.}]
			\step[fieldsource=\regexp{$MAPLOOP}, match={Biomedical}, replace={Biomed.}]
			\step[fieldsource=\regexp{$MAPLOOP}, match={Communications}, replace={Commun.}]
			\step[fieldsource=\regexp{$MAPLOOP}, match={Communication}, replace={Commun.}]
			\step[fieldsource=\regexp{$MAPLOOP}, match={Biophysics}, replace={Biophys.}]
			\step[fieldsource=\regexp{$MAPLOOP}, match={Compatibility}, replace={Compat.}]
			\step[fieldsource=\regexp{$MAPLOOP}, match={British}, replace={Brit.}]
			\step[fieldsource=\regexp{$MAPLOOP}, match={Components}, replace={Compon.}]
			\step[fieldsource=\regexp{$MAPLOOP}, match={Component}, replace={Compon.}]
			\step[fieldsource=\regexp{$MAPLOOP}, match={Computational}, replace={Comput.}]
			\step[fieldsource=\regexp{$MAPLOOP}, match={Delivery}, replace={Del.}]
			\step[fieldsource=\regexp{$MAPLOOP}, match={Computers}, replace={Comput.}]
			\step[fieldsource=\regexp{$MAPLOOP}, match={Computer}, replace={Comput.}]
			\step[fieldsource=\regexp{$MAPLOOP}, match={Department}, replace={Dept.}]
			\step[fieldsource=\regexp{$MAPLOOP}, match={Computing}, replace={Comput.}]
			\step[fieldsource=\regexp{$MAPLOOP}, match={Design}, replace={Des.}]
			\step[fieldsource=\regexp{$MAPLOOP}, match={Condensed}, replace={Condens.}]
			\step[fieldsource=\regexp{$MAPLOOP}, match={Detector}, replace={Detect.}]
			\step[fieldsource=\regexp{$MAPLOOP}, match={Conferences}, replace={Conf.}]
			\step[fieldsource=\regexp{$MAPLOOP}, match={Conference}, replace={Conf.}]
			\step[fieldsource=\regexp{$MAPLOOP}, match={Development}, replace={Develop.}]
			\step[fieldsource=\regexp{$MAPLOOP}, match={Congress}, replace={Congr.}]
			\step[fieldsource=\regexp{$MAPLOOP}, match={Differential}, replace={Differ.}]
			\step[fieldsource=\regexp{$MAPLOOP}, match={Consumer}, replace={Consum.}]
			\step[fieldsource=\regexp{$MAPLOOP}, match={Digest}, replace={Dig.}] 
			\step[fieldsource=\regexp{$MAPLOOP}, match={{ of the }}, replace={{ }}]
			\step[fieldsource=\regexp{$MAPLOOP}, match={{ of }}, replace={{ }}]
			\step[fieldsource=\regexp{$MAPLOOP}, match={{Of }}, replace={{ }}]
			\step[fieldsource=\regexp{$MAPLOOP}, match={{ on }}, replace={{ }}]
			\step[fieldsource=\regexp{$MAPLOOP}, match={{ On }}, replace={{ }}]
			\step[fieldsource=\regexp{$MAPLOOP}, match={{On }}, replace={{ }}]
			\step[fieldsource=\regexp{$MAPLOOP}, match={{ in }}, replace={{ }}]
			\step[fieldsource=\regexp{$MAPLOOP}, match=\regexp{\\\x{26}}, replace={{and}}] 
			\step[fieldsource=\regexp{$MAPLOOP}, match={{, and }}, replace={{, }}]
			\step[fieldsource=\regexp{$MAPLOOP}, match={{, And }}, replace={{, }}]
			\step[fieldsource=\regexp{$MAPLOOP}, match={{ and }}, replace={{ }}]
			\step[fieldsource=\regexp{$MAPLOOP}, match={{ And }}, replace={{ }}]
			\step[fieldsource=\regexp{$MAPLOOP}, match={{ In }}, replace={{ }}]
			\step[fieldsource=\regexp{$MAPLOOP}, match={{In }}, replace={{ }}]
			\step[fieldsource=\regexp{$MAPLOOP}, match={{ the }}, replace={{ }}] 
			\step[fieldsource=\regexp{$MAPLOOP}, match={{ The }}, replace={{ }}] 
			\step[fieldsource=\regexp{$MAPLOOP}, match={{The }}, replace={}] 
			\step[fieldsource=\regexp{$MAPLOOP}, match={{ for }}, replace={{ }}] 
			\step[fieldsource=\regexp{$MAPLOOP}, match={First}, replace={1st}]
			\step[fieldsource=\regexp{$MAPLOOP}, match={Second}, replace={2nd}]
			\step[fieldsource=\regexp{$MAPLOOP}, match={Third}, replace={3rd}]
			\step[fieldsource=\regexp{$MAPLOOP}, match={Fourth}, replace={4th}]
			\step[fieldsource=\regexp{$MAPLOOP}, match={Fifth}, replace={5th}]
			\step[fieldsource=\regexp{$MAPLOOP}, match={Sixth}, replace={6th}]
			\step[fieldsource=\regexp{$MAPLOOP}, match={Seventh}, replace={7th}]
			\step[fieldsource=\regexp{$MAPLOOP}, match={Eighth}, replace={8th}]
			\step[fieldsource=\regexp{$MAPLOOP}, match={Ninth}, replace={9th}]
			\step[fieldsource=\regexp{$MAPLOOP}, match={Tenth}, replace={10th}]
			\step[fieldsource=\regexp{$MAPLOOP}, match={Eleventh}, replace={11th}]
			\step[fieldsource=\regexp{$MAPLOOP}, match={Twelfth}, replace={12th}]
			\step[fieldsource=\regexp{$MAPLOOP}, match={Thirteenth}, replace={13th}]
			\step[fieldsource=\regexp{$MAPLOOP}, match={Fourteenth}, replace={14th}]
			\step[fieldsource=\regexp{$MAPLOOP}, match={Fifteenth}, replace={15th}]
			\step[fieldsource=\regexp{$MAPLOOP}, match={Sixteenth}, replace={16th}]
			\step[fieldsource=\regexp{$MAPLOOP}, match={Seventeenth}, replace={17th}]
			\step[fieldsource=\regexp{$MAPLOOP}, match={Eighteenth}, replace={18th}]
			\step[fieldsource=\regexp{$MAPLOOP}, match={Nineteenth}, replace={19th}]
			\step[fieldsource=\regexp{$MAPLOOP}, match={Twentieth}, replace={20th}]
		}
	}
}

\makeatletter
\newcommand{\linebreakand}{%
  \end{@IEEEauthorhalign}
  \hfill\mbox{}\par
  \mbox{}\hfill\begin{@IEEEauthorhalign}
}
\makeatother

\title{Approaching Safety-Argumentation-by-Design: A Requirement-based Safety Argumentation Life Cycle for Automated Vehicles

\thanks{The foundation of this work was supported by the German Federal Ministry for Economic Affairs and Energy within the project ``Automatisierter Transport zwischen Logistikzentren auf Schnellstraßen im Level 4 (\mbox{ATLAS-L4}).'' Additionally, this contribution was fostered as part of a bilateral collaboration between TRATON R\&D Germany GmbH and TU Braunschweig. The decisions, opinions, and content presented in this contribution reflect solely the views of the authors. This paper does neither represent the official position of TRATON, Volkswagen AG, or MOIA nor does it imply any specific development projects or strategic decisions by these companies.}
}

\author{\IEEEauthorblockN{%
    Marvin Loba\IEEEauthorrefmark{1},
    Robert Graubohm\IEEEauthorrefmark{1},
    Niklas Braun\IEEEauthorrefmark{1},
    Nayel Fabian Salem\IEEEauthorrefmark{1},\\
    Andreas Dotzler\IEEEauthorrefmark{2},
    Marcus Nolte\IEEEauthorrefmark{3},
    Torben Stolte\IEEEauthorrefmark{4},
    Richard Schubert\IEEEauthorrefmark{1},
    Markus Maurer\IEEEauthorrefmark{1}
}\\ 
\linebreakand
\IEEEauthorblockA{%
    \IEEEauthorrefmark{1}\textit{TU Braunschweig} \\%
    \textit{Institute of Control Engineering}, Braunschweig, Germany\\%
    \{m.loba, r..graubohm, niklas.braun, n.salem, t.stolte, r.schubert, markus.maurer\}@tu-braunschweig.de%
}
\linebreakand
\linebreakand
\linebreakand

\IEEEauthorblockA{%
    \IEEEauthorrefmark{2}\textit{TRATON R\&D Germany GmbH} \\
    Munich, Germany\\%
    andreas.dotzler@man.eu\\%
}

\and
\IEEEauthorblockA{%
    \IEEEauthorrefmark{3}%
     \textit{KTH Royal Institute of Technology}\\%
     \textit{Department of Engineering Design} \\
     \textit{Unit Mechatronics}, Stockholm, Sweden \\
    mnolte@kth.se%
}

\and 
\IEEEauthorblockA{%
    \IEEEauthorrefmark{4}%
    \textit{MOIA GmbH} \\
    \textit{Berlin, Germany}\\%
    torben.stolte@moia.io
}

}

\begin{document}



\twocolumn[
\begin{@twocolumnfalse}
\Huge {IEEE copyright notice} \\ \\
\large {\copyright\ 2026 IEEE. Personal use of this material is permitted. Permission from IEEE must be obtained for all other uses, in any current or future media, including reprinting/republishing this material for advertising or promotional purposes, creating new collective works, for resale or redistribution to servers or lists, or reuse of any copyrighted component of this work in other works.} \\ \\
{\Large Accepted to be published in \emph{2026 IEEE 29th International Conference on Intelligent
Transportation Systems (ITSC)}, Naples, Italy, September 15-18, 2026.}  \\ \\ 
Cite as:\\ \\
\noindent\fbox{%
    \parbox{\textwidth}{%
        M.~Loba, R.~Graubohm, N.~Braun, N.~F.~~Salem, A.~Dotzler, M.~Nolte, T.~Stolte, R.~Schubert, and M.~Maurer, ``Approaching Safety-Argumentation-by-Design: A Requirement-based Safety Argumentation Life Cycle for Automated Vehicles,'' in \emph{2026 IEEE 29th International Conference on Intelligent Transportation Systems (ITSC)}, Naples, Italy, September 15-18, 2026, {to be published}.
    }%
}
\vspace{2cm}

\end{@twocolumnfalse}
]

\noindent\begin{minipage}{\textwidth}

\hologo{BibTeX}:
\footnotesize
\begin{lstlisting}[frame=single]
@inproceedings{loba_SafetyArgLifeCycle_2026,
  author={Loba, M. and Graubohm, R. and Braun, N. and Salem, N. F. and Dotzler, A. and Nolte, M. and Stolte, T. and Schubert, R. and Maurer, M.},
  booktitle={2026 IEEE 29th International Conference on Intelligent Transportation Systems (ITSC)},
  title={Approaching Safety-Argumentation-by-Design: A Requirement-based Safety Argumentation Life Cycle for Automated Vehicles},
  address={Naples, Italy},
  year={2026},
  publisher={IEEE, to be published}
}
\end{lstlisting}
\end{minipage}

\addtolength{\topmargin}{46pt}

\maketitle

\addtolength{\topmargin}{-32pt}


\begin{abstract}

Despite the growing number of automated vehicles on public roads, operating such systems in open contexts inevitably involves incidents. 
Developing a defensible case that the residual risk is reduced to a reasonable (societally acceptable) level is hence a prerequisite to be prepared for potential liability cases. 
A ``safety argumentation'' is a common means to represent this case.
In this paper, we contribute to the state of the art in terms of process guidance on argumentation creation and maintenance---aiming to promote a \emph{safety-argumentation-by-design} paradigm, which mandates co-developing both the system and argumentation from the earliest stages.
Initially, we extend a systematic design model for automated driving functions with an argumentation layer to address prevailing misconceptions regarding the development of safety arguments in a process context.
Identified limitations of this extension motivate our complementary design of a dedicated argumentation life cycle that serves as an additional process viewpoint. 
Correspondingly, we define literature- and expert-based process requirements.
To illustrate the safety argumentation life cycle that we propose as a result of implementing these consolidated requirements, we demonstrate principles of the introduced process phases (\textit{baselining}, \textit{evolution}, \textit{continuous maintenance}) by an argumentation example on an operational design domain exit response.

\end{abstract}


\section{Introduction}
\label{sec:intro}


With the growing presence of automated driving applications on public roads, addressing the risk emerging from the vehicles' operation is crucial.
Due to an inherent risk in road traffic, which can be attributed to functional and systemic uncertainties~\cite{maurer2018}, the occurrence of safety-relevant incidents is inevitable when automated vehicles operate in a real-world environment, i.e., an open context\footnote{Refers to an environment that cannot be fully specified at design time, either due to its complexity, unpredictability, or temporal development~\cite{BurtonHawkins_2020}.}.
In anticipation of future incidents, the question of legal consequences immediately arises.
Decision makers in organizations need to be equipped with means to avoid personal and company liability.
While \emph{safety} linguistically is an ``open signifier''~\cite{Fleischer2023} and thus subject to implicitly deviating stakeholder understandings~\cite{Salem2026}, its definition established in the automotive domain is the \textit{absence of unreasonable risk}~\cite{ISO26262_2018,ISO21448_2022}.
As a consequence, prior to the product release as well as post-deployment, manufacturers need to provide a convincing reasoning that automated vehicle operations will not pose unreasonable risk---a reasoning that will be scrutinized again in the event of liability cases.

There is a substantial increase in safety assurance complexity when transitioning from proven driver assistance to automated driving systems.
Former practices of making a case for safety by simply consolidating normative work products are no longer sufficient to establish a release basis and demonstrate due diligence.
This is exacerbated since the current normative landscape is highly dynamic and far from flawless~\cite{Nolte2025}.
Regulation likewise shows issues, e.g., regarding different notions of \emph{safety}~\cite{Nolte2025AFGBV}.
Conversely, a coherent argumentation is required that argues how evidence contributes to achieving \emph{safety}.

Developing a structured argument supported by a body of evidence is a common approach to make a defensible case for system safety, both in standardization~\cite{UL4600_2023,ISO5083_2025,iso_pas_8800_2024} and best practice approaches~\cite{WaymoSafetyCase26}.
Providing a ``safety argumentation'' is also required by international regulation to obtain type approval of automated vehicles~\cite{eu1426}.
This implies that crafting this artifact is state of the art and thereby reduces legal risk of a provable defect according to product liability law. 

A notation commonly used for structured argumentations is the Goal Structuring Notation (GSN~\cite{GSN_Standard_2021}).
Utilizing such semi-formal notation promotes coping with increasing assurance complexity.
This is due to the fact that an argumentation model utilizes concepts for complexity management like hierarchy or modularity. 
The benefits of a structured safety argument hence extend far beyond legal compliance. 
Instead, moving past a ``checklist'' mentality (in terms of solely claiming adherence to normative and regulatory requirements) allows to facilitate risk reduction and communication.

Practical guidance and an agreed-upon state of the art are decisive for enabling automated driving.
We deem four safety argumentation-specific state of the art dimensions especially relevant.
On the one hand, this refers to argumentation content, argumentation structure, and argumentation techniques/methods.
On the other hand, this relates to processes for developing an argumentation.
This paper aims to narrow a gap in the latter dimension, i.e., the state of the art on processes dealing with safety argumentations for automated vehicles, which we find to be underrepresented in published literature.

Following a brief overview of our working terminology (\Cref{sec:terminology}), we present four interrelated contributions:
\begin{itemize}
    \item To counteract misconceptions regarding safety argumentations in development processes, we extend a process ($\blacktriangleright$~\mbox{\Cref{sec:ArgumentbyDesign}}) for the systematic design of automated driving functions with an argumentation layer to promote a \emph{safety-argumentation-by-design} paradigm.
    \item  Building on the need for an argumentation-focused viewpoint, we elicit requirements ($\blacktriangleright$~\Cref{sec:Requirements}) for designing a dedicated safety argumentation life cycle, complementing our previously introduced process extension.
    \item We propose the ``safety argumentation life cycle'' by implementing the specified requirements~($\blacktriangleright$~\Cref{sec:process}). 
    \item We use a case study to illustrate applying the safety argumentation life cycle's principles (e.g., \textit{evolution}) to a GSN-based argumentation on ``safe'' operation at operational design domain (ODD) limits ($\blacktriangleright$~\Cref{sec:casestudy}).
\end{itemize}



\section{TERMINOLOGY}
\label{sec:terminology}

\quad In the remainder of this paper, we consistently

\begin{enumerate}
    \item prefer ``\textbf{safety argumentation}'' over ``safety case,''\footnote{While the terms are oftentimes used interchangeably, we consider this distinction helpful to provide clarity on the artifact in question. 
    In particular, the safety case comprises the safety argumentation as well as the documented evidence and context that are referenced within the (for example GSN-based) argumentation.
    The argumentation depth of the GSN model, ultimately, poses a design decision influencing the necessary fragmentation of evidence.
    An ontology supporting the terminological delimitation can be found in~\cite{loba_2025}.}
    \item refer to the definition of safety established in automotive engineering as \textbf{``absence of unreasonable risk''}~\cite{ISO26262_2018},
    \item use \textbf{``safety argumentation life cycle''} when referring to the introduced process that captures the creation and maintenance of a safety argumentation, and
    \item use the term ``\textbf{system life cycle [process]}'' instead of ``development process'' to emphasize considering activities related to both development and operation.
\end{enumerate}


\section{Safety-Argumentation-by-Design}
\label{sec:ArgumentbyDesign}

In this section, first, we reflect on conflicting understandings when it comes to illustrating safety argumentations within process models for developing complex systems (\Cref{subsec:OpposingViewsKelly}).
Subsequently, we respond to identified potentials in terms of streamlining depictions for clarity.
To this end, we build upon a domain-specific design/development process for automated driving functions by adding an iterative layer that accounts for the evolution of a safety argumentation.
Discussing the current limitations of this extended process model
allows us to motivate both future work in this context but particularly the complementary design of a dedicated argumentation life cycle that we present in~\Cref{sec:Requirements,sec:process,sec:casestudy}.

\subsection{Views on Safety Argumentations in Development Processes}
\label{subsec:OpposingViewsKelly}

\begin{figure}[!h]
    \centering
    \includegraphics[width=\columnwidth]{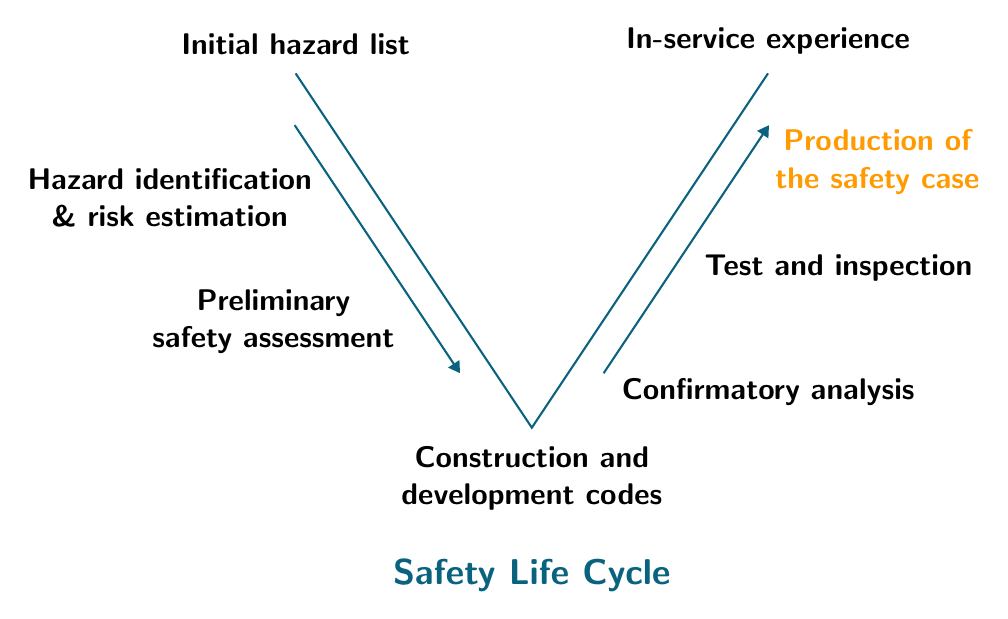}
	\caption[]{``Historical view of safety case development,'' adopted from~\cite[]{kelly2018safetycases}   
    }
	\label{fig:KellyPerspectivesHistorical} 
\end{figure}

In the context of safety-critical industries, Kelly~\cite{kelly2018safetycases} illustrates opposing perspectives on the production of a safety argumentation.
On the one hand, there is the historical way where the production of the safety argumentation is not performed until all other design and safety life cycle activities are conducted (\Cref{fig:KellyPerspectivesHistorical}).
As he explains, there are shortcomings to this traditionally practiced approach, e.g., extensive redesign of an argumentation.
Our experience is that such a ``wrapping up'' view is widespread in the automotive domain.
This observation coincides with remarks from Hobbs et al.~\cite{hobbs2023driving} who point out that the standard ISO~26262~\cite{ISO26262_2018} could be interpreted as handling the argumentation as ``something to complete just before deployment''~\cite{hobbs2023driving}.\footnote{While we perceive this view of consecutive ISO~26262 process stages (fostered by the V-Model illustration) prevails in the field, we acknowledge that ISO~26262, in principle, allows for iteratively revisiting previously conducted activities where necessary (see, e.g.,~\cite[Part~2,~5.4.6.1)]{ISO26262_2018}).}

\begin{figure}[!t]
    \centering
    \includegraphics[width=.7\columnwidth]{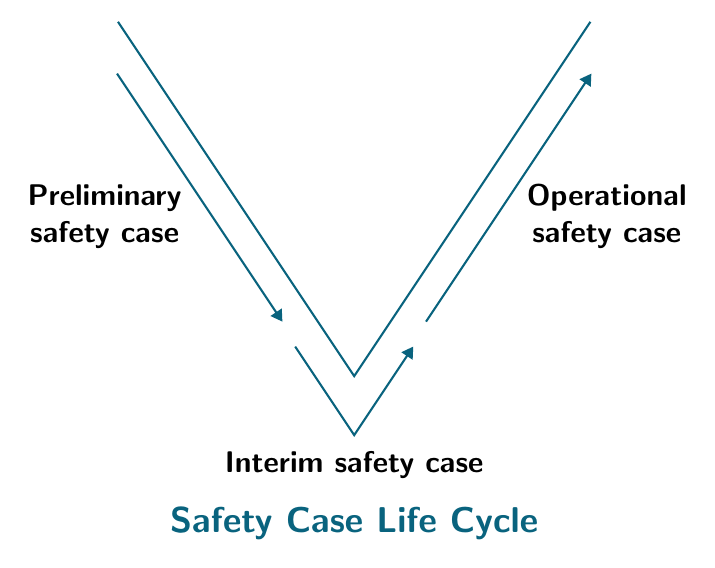}
	\caption[]{``Integrated view of safety case development,'' adopted from~\cite[]{kelly2018safetycases}   
    }
	\label{fig:KellyPerspectivesIntegrated} 
\end{figure}

On the other hand, Kelly~\cite{kelly2018safetycases} presents a favorable ``integrated'' view with distinct argumentation stages transitioning into each other over time.
Kelly highlights three stages of argumentation maturity (see~\Cref{fig:KellyPerspectivesIntegrated}).
These are specified to exist after defining and reviewing the system requirements specification (\textit{preliminary}), after initial system design and preliminary validation activities (\textit{interim}), and prior to in-service use including complete evidence of satisfying systems requirements (\textit{operational})~\cite{kelly2018safetycases}.

Hobbs et al.~\cite{hobbs2023driving} contrast the ``wrapping up'' perspective associated with ISO~26262 by claiming how the standard UL 4600~\cite{UL4600_2023} ``can be interpreted as building the product development on the safety case'' instead.
Associated with the disparity in these approaches, they reason how gathering evidence before formulating the argument structure results in an argumentation largely built around (early collected) available evidence---a bias that already led to severe incidents for safety-critical technology, for instance in aviation~\cite{haddencave2009nimrod}.

Strunk and Knight~\cite{Strunk2006} stress the existence of a bidirectional link between artifacts and an argumentation in the context of ``assurance-based development.''
This co-development paradigm is based on the premise that the argumentation and the system development activities to generate artifacts are tied closely.
Accordingly, the system and the argumentation should not only be developed in parallel but jointly.
This leads to synergy effects, since ``integration enables the whole system 
development process to benefit from the careful thought [...] that is required to create the system's assurance case''~\cite{Strunk2006}.

To conclude, we want to prevent the understanding that a sound safety argumentation can result from a mere consolidation of normative work products.
While the amount of documented evidence will naturally increase over time, a sufficiently valid argumentation will not simply arise from accumulating it.
Instead, developing the argumentation needs to be an integral part to the system life cycle.
Conversely, systematic thought and ongoing effort invested to craft the argumentation actively impacts development activities, hence posing a prerequisite to claim safety prior to deployment.

\subsection{Argumentation \& System Co-development Paradigm}
\label{subsec:ProcessExtension}

\begin{figure*}[!t]
    \centering
    \includegraphics[width=0.78\textwidth]{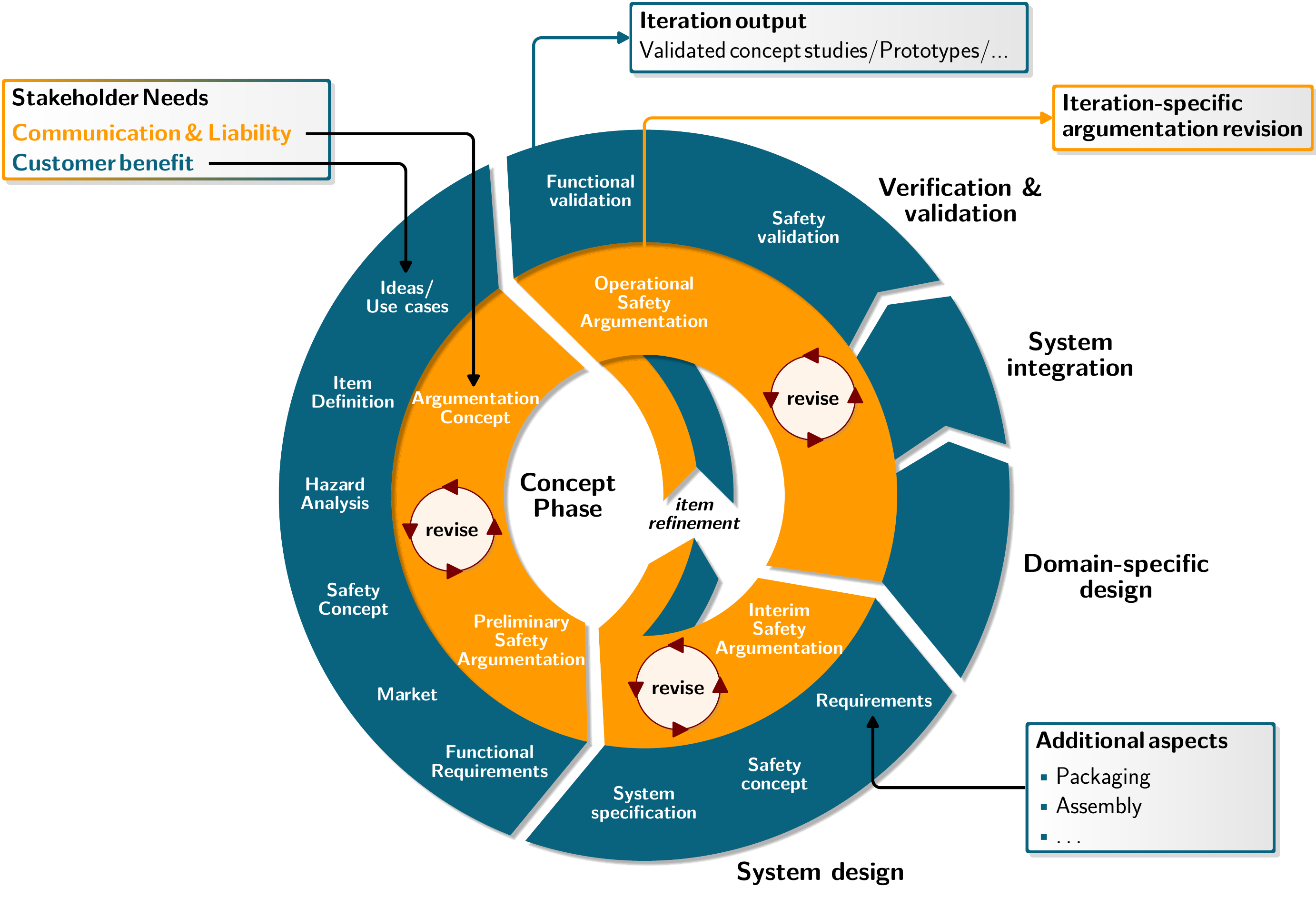}
	\caption[]{Extension of the reference process for the systematic design of automated driving functions (depicted in blue) explained in~\cite{winner_besondere_2026}.
    The added argumentation layer (red) runs in parallel, allocating different stages of argumentation maturity that are discussed by Kelly~\cite{kelly2018safetycases} and revisited with every (outer) iteration.
    Further additions are the incorporation of stakeholder needs in the concept phase, iteration-specific argumentation revisions as output, as well as \emph{revise} loops within phases that indicate ongoing revision of the evolving safety argumentation.
    Process phases are distinguished via \LegendBlueLoop{blue segments} and \LegendOrangeLoop{orange segments}.}
	\label{fig:ProcessExtension} 
\end{figure*}

We assess Kelly's application of the V-Model as a primarily illustrative choice, intended to build upon a widely recognized framework---aiming to convey the idea of how a safety argumentation should be understood to gradually develop throughout design and development while exhibiting increasing levels of maturity in an intuitive way.  
Still, the V-Model has inherent limitations when using it for visualization in the context of systems that entail significant complexity.
This is especially due to the fact that it can foster the perception of known requirements as input to development as well as a general lack of indicating the agile nature of complex system design.

As a basis for a domain-specific process model, these aspects are discussed in detail in~\cite{winner_besondere_2026}. 
The resulting model for the systematic design of automated driving functions is an example of a highly iterative process.
Correspondingly, the blue colored elements depicted in~\Cref{fig:ProcessExtension} are adopted from~\cite{winner_besondere_2026}.
This encompasses the outer loop that leads to iteration outputs of increasing maturity, the (optional) inner loop of \emph{item refinement} that triggers adapting functional objectives in case of an unsatisfactory safety concept/infeasible requirements---as well as associated process activities (e.g.,~\emph{safety validation}), resulting evidence (e.g.~\emph{item definition}), and relevant design inputs (e.g.,~\emph{additional aspects}).

To respond to the previously discussed necessity of a co-development paradigm, which reflects in a continuous refinement of the safety argumentation along the system life cycle, we do not simply allocate the argumentation as artifact/activity to be revisited with new iterations.
Instead, as indicated in red, \Cref{fig:ProcessExtension} shows our extension of the original systematic design model (blue) by an integrated argumentation layer.
We map the distinct argumentation stages introduced by Kelly (see~\Cref{fig:KellyPerspectivesIntegrated}) to specific process phases.
In particular, we highlight an ongoing argumentation evolution (depicted as \emph{revise} loops).

Similar to the customer benefit, which poses a motivation feeding the definition of ideas and use cases (upper left of~\cref{fig:ProcessExtension}), we deem the avoidance of \emph{liability}---as discussed in the introduction---a key stakeholder need that motivates the need for a coherent safety argumentation to make a defensible case for sufficient risk reduction in face of incidents.
Furthermore, in line with Homann~\cite{homann_wirtschaft_2005}, we deem it decisive that (residual) risk of safety-critical technology can and will be communicated from an early phase on.
On the one hand, we generally consider a safety argumentation to be a representation suitable to facilitate (internal and external) communication. 
On the other hand, this makes scrutinizing the arguability of safety as early as possible mandatory.
Thus, we add an \emph{argumentation concept} that needs to be specified at the ``start'' of each iteration (i.e., in the concept phase already).

The feedback loops account for development uncertainty and the emergence of design conflicts due to offering early iterations during the concept phase.
Consequently, this yields an approach that relies on extensive revisiting of activities to incrementally ``build safety'' into the system.
Driving early safety analyses in the sense of a \emph{safety-by-design} contrasts with the expectancy that safety can be established simply by scaling the testing effort post development.
The relevance of such paradigms is also evident in standards for automated vehicles~\cite{ISO5083_2025}.
By extending the systematic design model presented in~\cite{winner_besondere_2026} (blue elements) with the argumentation-specific inner layer (red elements), we visually emphasize the system and argumentation co-development paradigm---stressing the need to initiate argumentation-specific activities at the earliest stage possible and, thus, advancing a \textit{safety-argumentation-by-design} paradigm.

Nevertheless, we recognize two limitations present with the process model:
While we plan to incorporate operation phase activities in future work, the systematic design model used for the process extension does not yet account for it.
Additionally, the \emph{revise} actions are still abstracted, leaving the underlying principles of the associated continuous argumentation evolution subject to interpretation.
Hence, we consider it imperative to establish a complementary ``viewpoint'' by proposing a dedicated safety argumentation life cycle that captures argumentation-related activities and principles throughout design, development, and runtime.


\section{Process Requirement Elicitation}
\label{sec:Requirements}

First, we present (\Cref{subsec:RelWork}) and analyze (\Cref{subsec:ReqLiterature}) related work to define literature-based process requirements.
As our review of current literature reveals no existing argumentation-dedicated life cycle approaches in the field, this analysis includes examining process specifications that incorporate the argumentation as an artifact or its creation as an activity (in a manner comparable to our contribution in~\Cref{sec:ArgumentbyDesign} but mostly rudimentary).
While the explanations and depictions accompanying the process specifications do not specifically target an argumentation life cycle, we consult them in order to translate implicitly and explicitly described principles on the handling of safety argumentations into process requirements.
Subsequently, we supplement the literature-based with expert-based requirements (\Cref{subsec:ReqSupp}).


\subsection{Related Work}
\label{subsec:RelWork}

Recent publications show a growing tendency to explicitly account for safety argumentations in process depictions, both in technology-agnostic contexts (e.g.,~\cite{hawkins2025sace,hawkins2021amlas,kelly2018safetycases}) and in the domain of automated driving (e.g.,~\cite{iso_pas_8800_2024,chen2025scalable,kaiser2024agile}).

Hawkins et al. provide methodologies for safety assurance of autonomous systems ~\cite{hawkins2025sace} and for guidance on the assurance of machine learning usage specifically~\cite{hawkins2021amlas}.
The two corresponding processes aim to incrementally build a safety argumentation based on GSN patterns whose instantiation is driven by the output of each of the defined safety assurance activities. 
Zeller outlines a process for continuous safety assessment in software-intensive safety-critical systems in~\cite{zeller2021continuous}.
He positions the crafting of a safety argumentation from collected evidence as final ``development step'' before certifying and releasing the respective system.
Kaiser et al. present an agile approach to safety argumentation for automated vehicles through model-based engineering and simulation in~\cite{kaiser2024agile}.
The normative document ISO~PAS~8800 guides the development of a structured argumentation for artificial intelligence (AI) systems used in road vehicles, illustrated as an activity spanning the entire system life cycle~\cite[Clause 8]{iso_pas_8800_2024}.
Likewise, the ``assurance case development'' that Chen et al. introduce alongside a safety assurance framework for automated vehicles, starts during the requirement engineering phase already and lasts beyond post-deployment~\cite{chen2025scalable}.
Nouri et al.~\cite{nouri2025devsafeops} embed the safety argumentation for automated driving systems within a process depiction for continuous development and operations, referred to as ``\mbox{DevSafeOps}'' to emphasize the integration of safety.

While we acknowledge how Kelly's work (\cite{kelly2004systematic,kelly2018safetycases}) captures desired process features by a structured argumentation life cycle, we are not aware of publications on the systematic (requirement-based) design of a phased life cycle for developing/maintaining safety argumentations of automated vehicles.


\subsection{Literature-based Requirement Elicitation}
\label{subsec:ReqLiterature}

Subsequently, we gradually define following requirements:

\begin{tcolorbox}[colback=white, colframe=BlauDark, sharp corners=southwest, rounded corners=southeast, boxrule=0.2mm, left=1mm, right=1mm, top=1mm, bottom=1mm,title=Literature-based requirements,fonttitle=\scshape]
	A dedicated process for the argumentation life cycle shall...
	\begin{itemize}
		\setlength{\itemsep}{.1em}
        
        \item[\ReqUnchecked] indicate \textbf{\textcolor{BlauDark}{evolution}} during design/development.\MacroReq
        \item[\ReqUnchecked] illustrate \textbf{\textcolor{BlauDark}{growing maturity}} through revisions.
        \MacroReq
        \item[\ReqUnchecked] indicate mechanisms to \textbf{\textcolor{BlauDark}{monitor}} and \textbf{\textcolor{BlauDark}{continuously update}} the argumentation post-deployment. \MacroReq
    
	\end{itemize}
\end{tcolorbox}




\RequirementLine{1) Evolutionary approach}{1}
Kelly illustrates an argumentation-centered life cycle perspective (\Cref{fig:KellyPerspectivesIntegrated}) in early publications already, clarifying that ``the presentation of an evolving safety argument'' poses the core of a staged argumentation production~\cite{kelly2004systematic}.
As he states, an evolutionary approach initiated at the earliest stage possible has also been required by, for instance, defense standards for decades~\cite{defstan_00-56_pt1_iss2}.
Still, despite authors acknowledging that argumentations are subject to evolution and increasing maturity, process depictions both within (e.g.,~\cite{kaiser2024agile}) and beyond (e.g.,~\cite{zeller2021continuous}) the domain can be interpreted to support the ``historical'' understanding, e.g., by placing the safety argumentation as a step in the upper right of a V-Model (similar to~\Cref{fig:KellyPerspectivesHistorical}).
To overcome aforementioned issues, we require the argumentation life cycle to explicitly indicate an ongoing argumentation evolution during design and development (\ReqText{1}).

\RequirementLine{2) Maturity through revisions}{2}
We deem the three distinct maturity stages (see~\Cref{fig:KellyPerspectivesIntegrated}) that we adopted from Kelly~\cite{kelly2004systematic,kelly2018safetycases} to create our process extension shown in~\Cref{fig:ProcessExtension}) helpful abstractions.
Still, we consider the illustration of a higher degree of granularity beneficial since the sheer number of changes to an evolving safety argumentation yields numerous revisions.
Hence, we require the argumentation's evolution to be characterized by a continuous increase in maturity achieved through applying structure-altering operations to the argumentation (\ReqText{2}).

\RequirementLine{3) Monitoring \& continuous improvement}{3}
Recently published literature both in normative~\cite{iso_pas_8800_2024} and research contexts~\cite{chen2025scalable} suggests handling the creation of a safety argumentation for automated vehicles as an activity spanning the design \emph{and} operation phase (supporting \ReqText{1}).
This turns the argumentation into an artifact not only directed backwards in stating that the previously designed system is safe, but also directed forwards in stating that the product will stay safe over future releases.
While Kelly does not directly depict this perspective in \Cref{fig:KellyPerspectivesIntegrated} (``operational safety case'' refers to pre-deployment), he separately discusses ``safety case maintenance''~\cite{kelly2018safetycases}.
In line with this, Swaminathan et al. outline a conceptualization of argumentation maturity stages, too---but further distinguish between a deployment-ready and a continuously updated version of the safety argumentation~\cite{swaminathan2025trl}.

From a system life cycle perspective, the need for establishing updating mechanisms to preserve the absence of unreasonable risk during field operation is a normative requirement (e.g., according to ISO 21448~\cite{ISO21448_2022}).
This corresponds to our previously motivated future work to further develop the ideas depicted by the process extension we presented in~\Cref{subsec:ProcessExtension} by adding an operation phase that includes activities such as \textit{release}, \textit{deploy}, \textit{collect data}, and \textit{update} (see also~\cite{ISOIECIEEE15288_2023,nouri2025devsafeops}).
From an argumentation life cycle perspective, ISO~PAS~8800 points out how certain operating conditions might affect validity of assurance arguments for AI systems in road vehicles~\cite[Clause~8.4~e)]{iso_pas_8800_2024}).
Additionally, concepts like safety performance indicators (SPIs) pose known means to monitor the validity of claims in safety argumentations~\cite{UL4600_2023}.
Thus, we require 
highlighting mechanisms that promote continuous argumentation maintenance after deployment (\ReqText{3}).




\subsection{Expert-based Supplementary Requirements}
\label{subsec:ReqSupp}

As discussed above, the field lacks a harmonized state of the art for developing safety argumentations for automated vehicles.
This is especially relevant due to the existence of various competing argumentation approaches that miss an evident reasoning what principles determine the argumentation structure.
Thus, we require the life cycle to presuppose an argumentation framework that is derived from the state of the art in terms of structuring safety argumentations (\ReqText{4}).
Such an argumentation ``baselining'' is, e.g., in line with~\cite{hawkins2025sace,buysse2025,loba_2025}, basically pursuing a GSN pattern-based approach. 

The previous discussion on evolution inherently conditions an increasing level of maturity and context.
Correspondingly, we require the safety argumentation life cycle to indicate a spectrum ranging from an context-open framework to a context-specific argumentation revision prior to deployment (\ReqText{5})---where a growing degree of context is established by altering the argumentation structure (\emph{revise} actions).

Finally, we require the safety argumentation life cycle to visually indicate framework updates due to gained knowledge that reveals baselining deficits (\ReqText{6}), yielding:

\begin{tcolorbox}[colback=white, colframe=BlauDark, sharp corners=southwest, rounded corners=southeast, boxrule=0.2mm, left=1mm, right=1mm, top=1mm, bottom=1mm,title=Expert-based supplementary requirements ,fonttitle=\scshape]
	A dedicated process for the argumentation life cycle shall...
	\begin{itemize}
		\setlength{\itemsep}{.1em}
        \item[\ReqUnchecked] presuppose a \textbf{\textcolor{BlauDark}{baseline}} for the argumentation structure that is aligned with the state of the art.
        \MacroReq
        \item[\ReqUnchecked] depict how the progression through evolution is characterized by continuously \textbf{\textcolor{BlauDark}{increasing context}}.
        \MacroReq
        \item[\ReqUnchecked] foster \textbf{\textcolor{BlauDark}{continuous improvement}} of the argumentation framework due to gained knowledge.
        \MacroReq
	\end{itemize}
\end{tcolorbox}


\section{Safety Argumentation Life Cycle Proposal}
\label{sec:process}



\cref{fig:ArgLifecycle} shows the safety argumentation life cycle we propose based on implementing the requirements.
It comprises three phases we elaborate on in the next subsections, including comments on how the phases account for specified requirements.

\subsection{Baselining}
To account for \ReqText{4}, we consider an adequate ``baseline'' a prerequisite to developing sophisticated safety argumentations.
This can be realized by providing GSN patterns, accompanied by detailed guidance on their instantiation. 
Accordingly, we expect an argumentation framework in the \textit{baselining} stage.
We demand context-openness from such a framework to promote applicability to different systems and use cases.
The framework needs to follow state of the art and science principles to be a suitable starting point for evolving the argumentation.
An example for such a framework is the structural requirement-based pattern approach introduced in~\cite{loba_2025}.

\begin{figure}[!t]
    \centering
    \includegraphics[width=\columnwidth]{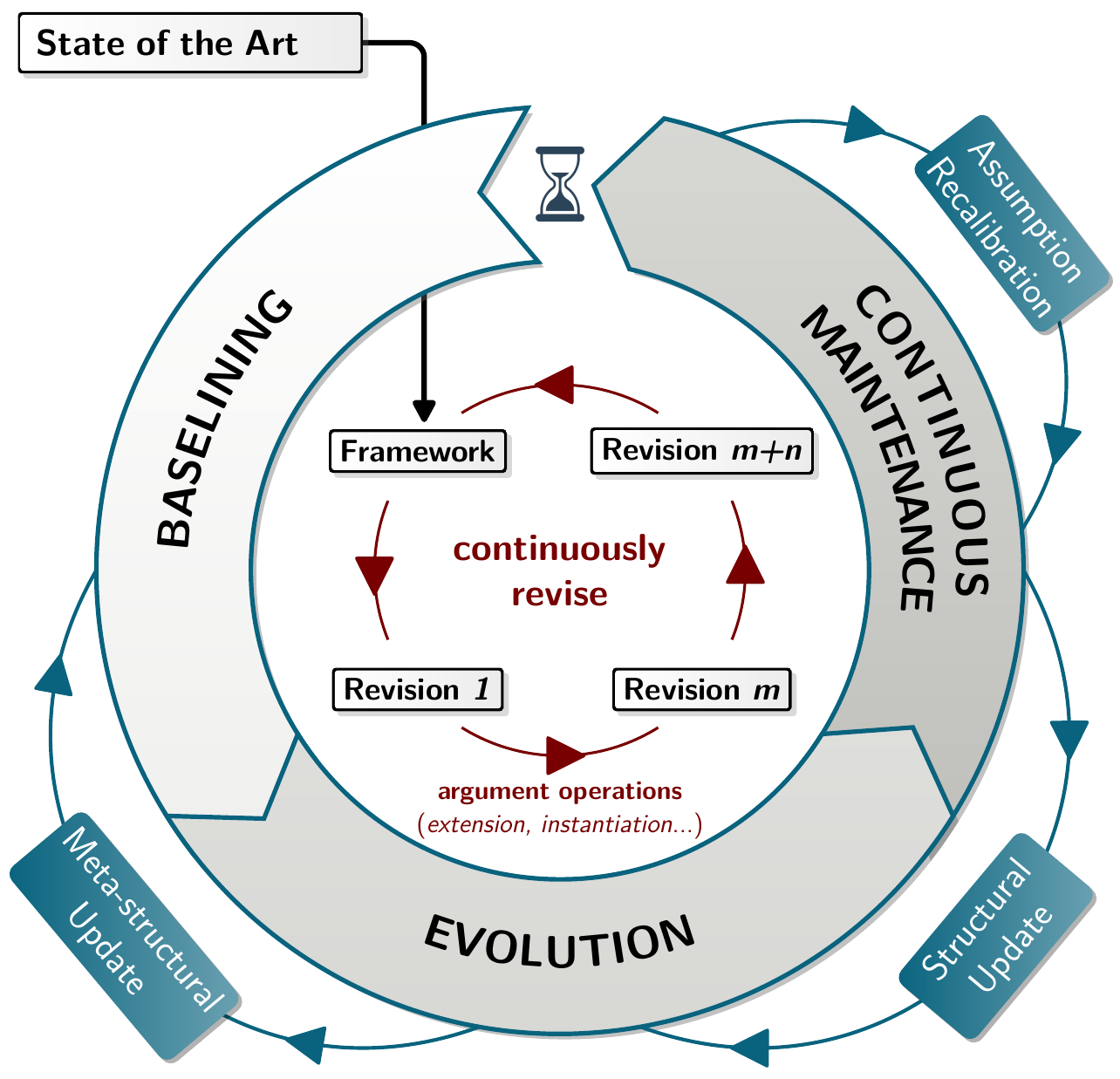}
	\caption[]{Resulting safety argumentation life cycle. 
    \LegContextFlow~indicates a flow of increasing maturity and context through revisions.
    \textcolor{darkRed}{\textit{Argument operations}} represent possible actions to adapt the argumentation structure.
    \LoopInline{PillBlue}{Updates} are illustrated by feedback loops applied within and between phases.
    }
	\label{fig:ArgLifecycle} 
\end{figure}

In contrast to the baseline, a ``deployment-ready'' argumentation revision $m$ eventually shall provide an argumentation tailored toward the specific context, i.e., the developed system of interest and its target operational domain.
The continuously increasing degree of context, e.g., due to design decisions taken, is visualized via a context flow (\ReqChecked{5}), spanning the whole life cycle and indicated by red arrows in~\Cref{fig:ArgLifecycle}.

In a strict sense, the baselining phase is not revisited after deployment (\emph{continuous maintenance}).
Still, the hour glass in~\Cref{fig:ArgLifecycle} depicts that activities associated with the baselining stage will need to be conducted again to appropriately adapt the framework for future projects and systems.


\subsection{Evolution}

The incremental growth in argumentation maturity is illustrated through revisions in the course of an ``evolution'' occurring along design and development time (\ReqChecked{1}~\&~\ReqChecked{2}).
Accordingly, a multitude of revisions results from different ``argument operations'' applied to each previous revision, where the first revision is inherited from the context-open framework.
These operations yield a continuous enrichment of the argumentation with context.
Example operations we propose (in a non-exhaustive manner) are \textit{Instantiation}, \textit{Elimination}, and \textit{Extension}.
The first relates to common principles of managing GSN patterns (see, e.g.,~\cite{hawkins2025sace,kelly1998arguing}).
Undeveloped GSN goals are developed by downstream sub-argumentations.
Placeholders used in GSN patterns are replaced by concrete instances.
Elimination and extension describe removing non-relevant argumentation parts or adding parts in case of evidently missing aspects.
We use the case study presented in \Cref{sec:casestudy} to illustrate applying these to a framework.

\subsection{Continuous Maintenance}
As previously discussed, Kelly expresses the idea of ongoing argumentation adaptation through the label \emph{maintenance}~\cite{kelly2018safetycases}.
This resembles principles of \emph{continuous} assurance in DevOps paradigms (see, e.g., ~\cite{nouri2025devsafeops,cassel2025devops}).
We want to clarify that at no point should there be an invalid argumentation post-deployment.
However, SPI violations can indicate necessary argumentation updates to prevent the emergence of unreasonable risk.
Correspondingly, Cassel et al. consider ``field evidence-based verification and validation as a first-class citizen''~\cite{cassel2025devops} when it comes to updates during operation.
Thus, field data collection plays a decisive role in ensuring life cycle validity of an automated vehicle's safety argumentation.
More precise, we consider the gathering of field evidence compensates for epistemological uncertainty and is required to ensure the operations remain ``safe.''

\Cref{fig:ArgLifecycle} shows how the \textit{m}-th revision is transferred and fed into the ``continuous maintenance'' phase when a system is deployed for commercial operation.
Threats towards assumption validity (e.g., SPI violations) for the argumentation are then monitored and acted upon when detected.
Complementarily, other triggers might inform improvements of the safety argumentation.
These can arise from the release of novel standards, laws, and best practices or new safety mechanisms that result from reacting to safety-relevant field incidents.
Still, the primary knowledge resource serving the improvement of a safety argumentation is collected field evidence.
The update mechanism are detailed in the next subsection (\ReqChecked{3}).

\subsection{Update Mechanisms \& Feedback Loops}

A concept meant to evaluate claim validity is the runtime implementation of SPIs~\cite{UL4600_2023}.
Detecting violations of defined SPI-associated thresholds invokes a root cause analysis on what led to the non-compliance.
Such an evaluation can yield different consequences intended to account for the identified validity threat.
In a phase of early commercialization, various individual SPI violations might occur, which do not necessarily result in adapting the argumentation structure---but mandate to recalibrate the initial assumptions underlying the specification of SPIs and/or the associated thresholds.
Certainly, other update-inducing causes, such as qualitative assessments, are likely to become relevant, too. 
This (predominantly SPI-based) \textit{assumption recalibration} is depicted as feedback mechanism within the continuous maintenance phase in~\Cref{fig:ArgLifecycle}.

A second (outer) feedback loop is shown running from the continuous maintenance to the evolution stage.
This refers to \textit{structural updates} that consist in altering the argumentation structure to counteract either the SPI-based detection of threats to assumption validity or respond to gained knowledge that triggers an improvement of the argumentation.
This may lead to additional \textit{n} revisions that again need to be monitored.

Another phase-overarching update mechanism affects the baselining.
As developing an argumentation framework is always subject to uncertainty, knowledge won over time will probably reveal deficits.
This information shall drive the incremental refinement of the underlying argumentation baseline.
A structural update that is triggered in the continuous maintenance stage may propagate to an update of the underlying argumentation framework, i.e., cause a \textit{meta-structural update}.
Alternatively, such a meta-structural update, depicted as outer feedback loop directed towards the baselining phase (\ReqChecked{7}), might become necessary during the evolution already.


\section{Case Study: ODD Exit Detection \& Response}
\label{sec:casestudy}

We use a minimal example of a GSN-based argument that builds on an underlying argumentation framework in order to demonstrate the concepts and principles we previously presented for the introduced safety argumentation life cycle.
More specifically, we use the example to demonstrate the application of selected argument operations (\textit{extension} and \textit{instantiation}) and SPI-based update mechanisms.

\begin{figure}[t!]
    \centering
    \includegraphics[width=\columnwidth]{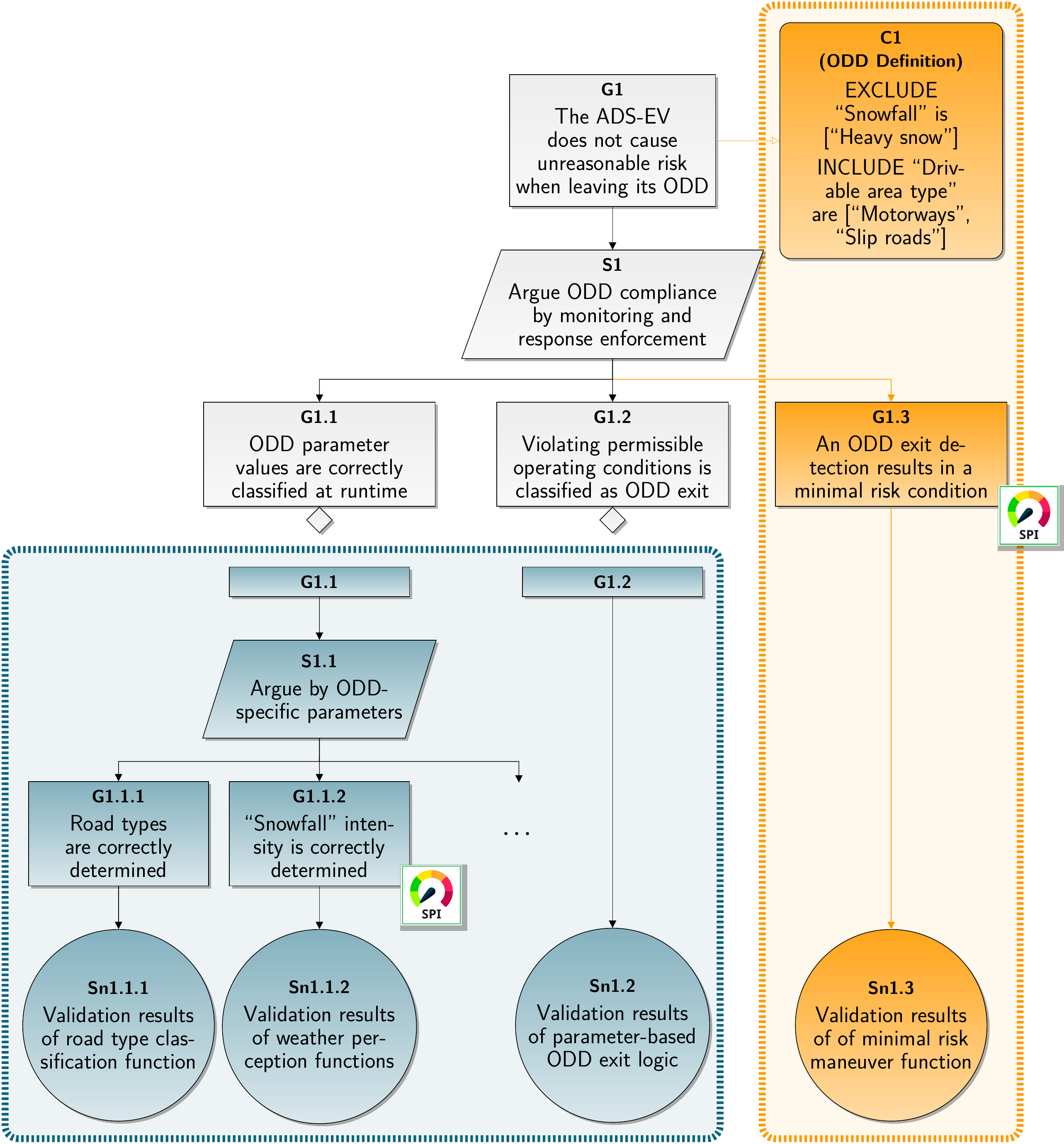}
	\caption[]{Example GSN argument toward ODD exit response. 
    The color coding indicates \LegendFramework, \LegendInstantiated, and~\LegendExtended GSN elements, respectively.
    \LegendUndeveloped represents an undeveloped claim.}
	\label{fig:CaseStudyGSN} 
\end{figure}

In~\Cref{fig:CaseStudyGSN}, we argue toward the claim G1 that a vehicle equipped with an automated driving system does not cause unreasonable risk when leaving its ODD.
While preventing operation outside of the ODD needs to be a core development objective, ODD exits are inevitable when operating in an open context.
In this case, the vehicle needs to achieve a \textit{minimal risk condition}~\cite{ISO22736_2021}, as the system is not designed to handle the encountered operating conditions.
An appropriate system response is commonly referred to as \textit{minimal risk maneuver}.
All gray elements are associated with an underlying argumentation baseline, i.e., supposed to be a specific part of an argumentation framework. 
The goal G1 is decomposed in order to argue over ODD compliance by runtime monitoring and response enforcement (GSN strategy S1).
The corresponding sub-goals claim the correct classification of ODD parameter values (G1.1) and the actual action of classifying an ODD exit as outcome of the evaluation of permissible operating conditions against conditions observed at runtime (G1.2).
The sub-claims are marked as ``undeveloped'' via the GSN diamond decorators below the elements, in line with the GSN pattern character of a context-open framework.

The argument adaptations that result as a consequence of applying the argument operations during the evolution stage are indicated by color (\textit{instantiation} in blue and \textit{extension} in orange).
Regarding the extension, two aspects are added in comparison to the argumentation framework.
These additions are based on the specific context of the system and operation of interest.
In this example, we assume a use case of Hub-to-Hub operation of automated trucks.
Hence, the added context element C1 provides an excerpt of our use case's ODD definition, formulated as compositional statement according to state of the art ODD taxonomy, ODD attributes, and ODD definition language syntax~\Cite{iso_34503_2023,bsi_pas_1883_2025}.
Accordingly, we exclude heavy snowfall and include motorways and slip roads as drivable area types, i.e., as permissible operating conditions.

Another extension is adding the sub-goal G1.3 that claims how an event classified as ODD exit (argued over in terms of ODD monitoring via G1.1 and G1.2) actually translates into the minimal risk maneuver execution.
Complementarily, the GSN solution Sn1.3 is added to support the fulfillment of G1.3 with respective evidence.
On first glance, it may seem counterintuitive that G1.3 would not be part of the original framework's argumentation pattern.
However, a comprehensive framework will always be subject to flaws.
In turn, the identified absence of this necessary claim would not only lead to applying an extension---but suggest a deficiency in the rigor applied during the framework's development phase and necessitate improving the framework (\textit{meta-structural update}).

The instantiation of G1.1 and G1.2 demonstrates how increased argumentation context might result from adding an evidence to be provided (Sn1.2) or supplementing the originally undeveloped claim G1.1 with a more detailed downstream argumentation.
Correspondingly, a strategy to argue over relevant ODD-specific parameters (S1.1) is utilized to decompose the undeveloped goal G1.1 into sub-claims, which we argue in terms of correct determination of the ODD attributes ``road types'' (G1.1.1) and ``snowfall'' (G1.1.2).

An SPI example suitable to monitor the validity of the claim G1.1.2 in the continuous maintenance phase might be a false negative rate for snowfall detection.
Consequences of such an SPI violation could be the recalibration of assumptions, e.g., ``just'' influencing the rate threshold definition.
Yet, a detection rate too low with respect to the ``correct'' determination of heavy snowfall could have a direct impact on the level of residual risk.
The system could encounter weather conditions the system is not designed to tolerate but cannot identify this as an event requiring an immediate activation of a minimal risk maneuver.
Consequently, there might be functional modifications, e.g., constraining the ODD or changing the sensor setup to enhance environmental perception---implying adaptions to the argumentation in the manner of a \emph{structural update}.
Another SPI could be a ``success rate'' of minimal risk maneuver executions associated with G1.3.
This means determining how many times a triggered minimal risk maneuver actually led to achieving a minimal risk condition.
An associated threshold violation could necessitate improving the minimal risk maneuver function.

\section{CONCLUSION AND FUTURE WORK}
\label{sec:Conclusion}


In this paper, we promote a \emph{safety-argumentation-by-design} paradigm that relies on driving the joint consideration of both system and safety argumentation development, starting as early as possible in the system lifecycle.
Hence, we extended a systematic design model for automated driving functions with an argumentation-specific layer to resolve process-related misconceptions on approaching the creation of safety argumentations. 
To provide sufficient depth, we complemented this extension with a dedicated safety argumentation life cycle that we designed based on previously defined requirements.
Finally, we demonstrated the fundamental concepts of the introduced lifecycle through a GSN case study on operating at ODD limits.
We are currently working on an article that builds upon this paper's proposal of a safety argumentation life cycle and particularly aims for the following linked contributions:

\begin{enumerate}
    \item Detailed presentation of a GSN-based state of the art argumentation framework for automated vehicles
    \item Proof of concept of the framework's \emph{evolution} toward a \emph{deployment-ready} argumentation for a concrete ODD, i.e., introducing a comprehensive example to enhance clarity on applying the argument operations
    \item Illustrating the concept of \emph{representation-supported communication} by elaborating the transformation of this \emph{deployment-ready} argumentation revision into a release document for a specific operational context
    \item Discussing organizational implications, such as emerging efforts (e.g., communication) and the necessity of novel roles, tasks, as well as stakeholder responsibilities
    
\end{enumerate}






\renewcommand*{\bibfont}{\footnotesize}

\printbibliography


\end{document}